\documentclass[preprint]{aastex}
\usepackage{emulateapj5}
\usepackage{apjfonts}
\usepackage{amssymb, amsmath}
\usepackage{graphicx}
\usepackage{CJK}
\usepackage{color}
\usepackage{xcolor}
\usepackage{capt-of}
\usepackage{float}
\DeclareGraphicsExtensions{.pdf,.png,.jpg}

\newdimen\digitwidth    
\setbox1=\hbox{0}       
\digitwidth=\wd1        
\catcode`"=\active      

\def"{\kern\digitwidth}

\def\aj{{AJ}}

\def\apj{{ApJ}}
\def\apjs{{ApJS}}
\def\asec{$^{\prime\prime}$}
\def\deg{$^{\circ}$}

\def\etal{{\ et al.\ }}

\def\lax{{$\mathrel{\hbox{\rlap{\hbox{\lower4pt\hbox{$\sim$}}}\hbox{$<$}}}$}}
\def\gax{{$\mathrel{\hbox{\rlap{\hbox{\lower4pt\hbox{$\sim$}}}\hbox{$>$}}}$}}
\def\simlt{\lower.5ex\hbox{$\; \buildrel < \over \sim \;$}}
\def\simgt{\lower.5ex\hbox{$\; \buildrel > \over \sim \;$}}

\def\mnras{{MNRAS}}

\newcommand{\lt}{<}
\newcommand{\gt}{>}

\def\ser{{S\'{e}rsic\ }}

\slugcomment{\textit{The Astrophysical Journal}, Accepted}
\shorttitle{THE CARNEGIE-IRVINE GALAXY SURVEY\@. IV.}
\shortauthors{HUANG ET AL.}

\begin{document}

\begin{CJK*}{UTF8}{gbsn}

\title{The Carnegie-Irvine Galaxy Survey\@. IV\@. A Method to Determine the Average Mass 
    Ratio of Mergers That Built Massive Elliptical Galaxies}

\author{Song Huang (黄崧)\altaffilmark{1} Luis C. Ho\altaffilmark{2,3}, 
    Chien Y. Peng\altaffilmark{4}, Zhao-Yu Li (李兆聿)\altaffilmark{5}, and 
    Aaron J. Barth\altaffilmark{6} }
\date{}                                          

\altaffiltext{1}{Kavli Institute for the Physics and Mathematics of the
    Universe, Todai Institutes for Advanced Study, the University of Tokyo 
    (Kavli IPMU, WPI), Kashiwa 277-8583, Japan}

\altaffiltext{2}{Kavli Institute for Astronomy and Astrophysics, Peking 
    University, Beijing 100871, China}

\altaffiltext{3}{Department of Astronomy, School of Physics, Peking University, Beijing
    100871, China}

\altaffiltext{4}{Giant Magellan Telescope Organization, 251 South Lake Avenue, 
    Suite 300, Pasadena, CA 91101, USA}

\altaffiltext{5}{Key Laboratory for Research in Galaxies and Cosmology,
    Shanghai Astronomical Observatory, Chinese Academy of Sciences,
    80 Nandan Road, Shanghai 200030, China}

\altaffiltext{6}{Department of Physics and Astronomy, 4129 Frederick Reines 
    Hall, University of California, Irvine, CA 92697-4575, USA}


\begin{abstract}

  Many recent observations and numerical simulations suggest that nearby massive,
  early-type galaxies were formed through a ``two-phase'' process.  In the proposed second
  phase, the extended stellar envelope was accumulated through many dry mergers.  However,
  details of the past merger history of present-day ellipticals, such as the typical
  merger mass ratio, are difficult to constrain observationally.  Within the context and
  assumptions of the two-phase formation scenario, we propose a straightforward method,
  using photometric data alone,  to estimate the average mass ratio of mergers that
  contributed to the build-up of massive elliptical galaxies.  We study a sample of nearby
  massive elliptical galaxies selected from the Carnegie-Irvine Galaxy Survey, using
  two-dimensional analysis to decompose their light distribution into an inner, denser
  component plus an extended, outer envelope, each having a different optical color.  The
  combination of these two substructures accurately recovers the negative color gradient
  exhibited by the galaxy as whole.  The color difference between the two components
  ($\langle\Delta(B-V)\rangle \simeq 0.10$ mag; $\langle\Delta(B-R)\rangle \simeq 0.14$
  mag), based on the slope of the $M_{\ast}$-color relation for nearby early-type
  galaxies, can be translated into an estimate of the average mass ratio of the mergers.
  The rough estimate, 1:5 to 1:10, is consistent with the expectation of the two-phase
  formation scenario, suggesting that minor mergers were largely responsible for building
  up to the outer stellar envelope of present-day massive ellipticals.  With the help of
  accurate photometry, large sample size, and more choices of colors promised by ongoing
  and future surveys, the approach proposed here can reveal more insights into the growth
  of massive galaxies during the last few Gyr. 

\end{abstract}

\maketitle


\section{Introduction}

\subsection{Two-phase Formation Scenario for Massive Galaxies}

    Recent observations established that massive, quiescent early-type galaxies (ETGs) at
    high redshift are surprisingly more compact (Daddi\etal 2005; Trujillo\etal 2006) than
    their descendants today.  Since $z=2$, these systems on average doubled in stellar
    mass and increased in size by a factor of 3--5 (Szomoru\etal 2012; van~der~Wel\etal
    2014).  A ``two-phase'' scenario (Oser\etal 2010) emerged as a plausible explanation
    for this dramatic evolution: intense dissipative processes rapidly built up an
    initially compact progenitor; then, after star formation was quenched, gas-poor
    (``dry''), minor mergers\footnote{The distinction between major and minor merger in
    terms of mass ratio is quite arbitrary.  The threshold most commonly adopted in the
    literature is a mass ratio of 1:3, which we also use here.} dominated the second phase
    of protracted evolution.  The inside-out growth of massive galaxies predicted by this
    picture has gained much recent observational support (e.g.,~Patel\etal 2013;
    Huang\etal 2013b).  The key feature of this scenario is the dominant role of minor,
    mostly dry, mergers in the build-up of the extended envelope, the feature most
    responsible for the dramatic size growth, of massive elliptical galaxies.  Many recent
    simulations have confirmed this and also predicted either $M_{\ast}$-weighted or
    number-weighted mass ratio for mergers during the second phase.  Using re-simulations
    of 40 massive halos, Oser\etal (2012) concluded that ``minor mergers with a
    mass-weighted mass ratio of 1:5'' dominate the accretion events at $z \lt 2$.  Tracing
    the evolution of 611 massive halos in an adaptive mesh-refinement simulation,
    Lackner\etal (2012) also concluded that accretion onto massive halos are not dominated
    by major mergers, even though the fraction of the accreted component estimated by
    these authors is quite different from that of Oser\etal (2012).  Lackner\etal also
    suggested that the median mass-weighted mass ratio should be smaller than 1:5.  A
    similar conclusion was reached by Gabor \& Dav{\'e} (2012): the mass growth of massive
    galaxies after they became quiescent was dominated by minor mergers with typical a
    mass ratio of 1:5.  

    Using N-body simulations of dry major and minor mergers, Hilz\etal (2012) showed that
    minor mergers can be very efficient in increasing the size once the dark matter halos
    are considered.  Later, Hilz\etal (2013) demonstrated that a few mergers with mass
    ratios of 1:5 can explain the observed size evolution of massive ETGs.  Oogi \& Habe
    (2013) performed ``sequential minor mergers with parabolic and head-on orbits,''
    including the dark matter halo for the satellites.  Minor mergers with mass ratios
    between 1:10 and 1:20 are able to reproduce the observed growth in size and change of
    mass profile.  The dry minor merger hypothesis was also tested by B{\'e}dorf \&
    Portegies Zwart (2013), who, again, favored minor mergers with mass ratios between 
    1:5 to 1:10.

    \vskip 0.3cm
    \figurenum{1}
    \begin{figure*}[t]
    \centering 
    \includegraphics[width=14.5cm]{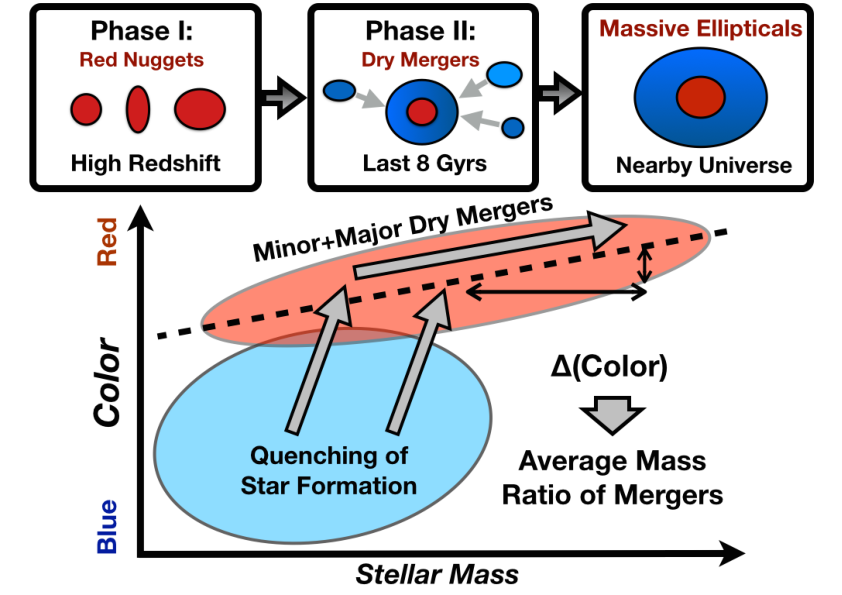}
    \caption{Cartoons that illustrate the basic idea of the proposed method for the
        estimation of average merger mass ratio.  The style is adopted from Faber\etal
        (2007).  The upper panels show the basic picture of the two-phase formation
        scenario of massive ellipticals. The lower panel describes a general picture for
        the evolution of massive galaxies on the $M_{\ast}$-color plane.  After massive
        ETGs were quenched, they quickly moved to the red sequence and gradually evolved
        along it as they became redder due to the aging of the stellar population and more
        massive due to the accumulation of extended stellar envelope through many minor
        (dry) mergers.  Since the stars of smaller systems should have lower
        metallicity, the outer envelope naturally has a bluer color compared to the inner
        region.  The color difference between the inner region and the outer envelope thus
        reflects the average mass ratio of all mergers that contributed to the second
        phase of the evolution.  Larger color difference means more stellar material came
        from smaller galaxies, making minor merger more important.  
    }
    \label{figure:1}
    \end{figure*}
  
    Although various observations point to the same direction (e.g., Coccato\etal 2010,
    2013; Greene\etal 2013; Lee\etal 2013), it is not clear whether enough minor mergers
    during the last 8--10 Gyr can account for the growth in size and mass of ETGs (e.g.,
    Bluck\etal 2012; Newman\etal 2012; L{\'o}pez-Sanjuan\etal 2012; Ferreras\etal 2014;
    Ownsworth\etal 2014).  Evidence of minor mergers is plentiful in the local Universe
    (e.g., Tal\etal 2009; Gu\etal 2013), but they are challenging to observe at higher
    redshift and hard to use for quantifying the merger mass ratio.  Satellite statistics
    provide another approach to constrain merger mass ratios.  Ruiz\etal (2014) estimated
    the distribution of $M_{\ast}$ of satellites around nearby, massive galaxies in the
    Sloan Digital Sky Survey (SDSS) down to a mass ratio of 1:400, and concluded that the
    merger channel should be dominated by minor mergers with mass ratio 1:10.  However,
    this method is quite sensitive to uncertainties in the estimations of $M_{\ast}$ and
    the mass completeness of the sample.  Meanwhile, valuable fossil information embedded
    in the outskirts of nearby massive galaxies should also be explored to provide better
    observational constraints of their merging history.  Arnold\etal (2011) measured the
    metallicity gradient of globular clusters in NGC~3115, and suggested its halo was
    formed through accretions of small objects with 15:1 to 20:1 mass ratio.  Greene\etal
    (2012, 2013, 2015) measured spatially resolved stellar population properties of $\sim
    100$ nearby ellipticals out to 2.5$R_{e}$.  The lower metallicity and high
    $\alpha$-element enhancement at large radii indicate a possible connection with
    smaller systems whose star formation was quenched at early time.  Although globular
    clusters are unique tracers of formation at early phase, and spectroscopic information
    at large radii is very promising in light of ambitious surveys such as MaNGA
    (Bundy\etal 2015), these observations are very expensive to acquire for a
    representative sample of galaxies.  Even with MaNGA, accurate stellar population
    properties are still hard to obtain for the faint outskirts of individual galaxies.
    And neither method can be applied easily to high redshifts.  

    Photometric colors also carry key information about the stellar population, and
    compared to the above-mentioned methods, are significantly easier to measure at large
    radii or high redshift.  In this paper, we will show that the average merger history
    of nearby massive galaxies can be constrained using the spatial distribution of their
    optical colors.
  
\subsection{Color Gradients of Massive Elliptical Galaxies} 

    It is well known that massive ETGs often show a negative gradient of optical color,
    with colors becoming redder toward smaller radii (de~Vaucouleurs 1961; Vader\etal
    1988; Franx \& Illingworth~1990; Peletier\etal 1990; see Kormendy \& Djorgovski 1989
    for a comprehensive review).  With the help of new observations, this important
    feature has been characterized using much larger samples, and observed at higher
    redshifts (Hinkley\etal 2001; Michard 2005; Ferreras\etal 2009; La Barbera\etal 2010;
    Tal \& van Dokkum 2011).  The negative gradient is usually considered to be caused by
    radial variation of stellar population, especially metallicity (e.g., Ogando\etal
    2005; Spolaor \etal 2010; Kuntschner\etal 2010; La~Barbera \etal 2010; Rawle\etal
    2010).  Numerous attempts have been made to explain color gradients within the
    formation scenario of massive galaxies.  Generally speaking, intense in-situ star
    formation, as predicted in early monolithic collapse models, tends to leave behind a
    steep, negative color gradient (e.g., Eggen\etal 1962; Larson 1975; Carlberg 1984;
    Kawata \& Gibson 2003; but also see Pipino\etal 2010), while major merger models
    (e.g., White 1978) predict significant flattening of the gradient (e.g., Kobayashi
    2004; Rupke\etal 2010).   

    Meanwhile, the two-phase formation scenario can naturally explain the negative color
    gradient: minor mergers tend to redistribute stars from the less massive system, which
    has lower stellar metallicity, to the outer envelope of the remnant (e.g., Hilz\etal
    2012), and help build up a bluer stellar halo (La~Barbera\etal 2013; D'Souza\etal
    2014).  Recently, Hirschmann\etal (2015) demonstrated that the color gradients of
    massive galaxies are largely due to the metallicity difference between the inner
    in-situ and outer accreted components (see their Figure~4).  Although the negative
    color gradient itself is no longer a new discovery, here we propose a new angle to
    interpret the color gradient with the help of our two-dimensional (2-D) modeling
    method and the $M_{\ast}$-color relation of ETGs.
    
\subsection{The Red Sequence and the $M_{\ast}$-color Relation}

    One of the most important scaling relations for ETGs is the tight correlation between
    their luminosity (or stellar mass) and their color (Sandage \& Visvanathan 1978).
    This red sequence was first noticed for ETGs in clusters, as it is very prominent in
    the color-magnitude plane, even without redshift information (e.g., Bower\etal 1992;
    Gladders\etal 1998).  Now, we know that it is a universal relation for all ETGs at
    $z\approx 0$, regardless of environment (e.g., Bernardi\etal 2003, 2005).  Furthermore,
    the red sequence has merged into the bigger picture of bimodality of galaxies on the
    $M_{\ast}$-color plane (Strateva\etal 2001; Baldry\etal 2004).  So far, the red
    sequence has been reliably detected up to $z \approx 1.5$ (e.g., Kodama\etal
    1998; Tanaka\etal 2005; Cassata\etal 2007; Mei\etal 2009; Nicol\etal 2011; Fritz\etal
    2014), and possibly higher (e.g., Kriek 2008; Whitaker\etal 2010).  The slope of the
    red sequence seems to not evolve much since $z \approx 1.0$ (e.g., Mei\etal 2009;
    Fritz\etal 2014).  The positive slope of the relation suggests that more massive ETGs
    tend to have higher average metallicity (also older luminosity-weighted age and higher
    $\alpha$-element enrichment; e.g., Faber 1973; Kodama\etal 1999; Bernardi\etal 2006;
    Graves\etal 2009).  Meanwhile, the evolution of the red sequence from $z \approx 1$ to
    0 is consistent with passive evolution of an old stellar population (e.g., Bell\etal
    2004; Fritz\etal 2014).  During the same period, the stellar mass on the red sequence
    has doubled (e.g., Faber\etal 2007).  Based on these observations, it would seem
    that massive ETGs most likely evolve along the red sequence during the last 8--9 Gyr
    after internal star formation has been quenched early and quickly, although the
    detailed evolutionary track (e.g., Faber\etal 2007; Schawinski\etal 2014) and
    mechanisms (e.g., Bernardi\etal 2011) still need to be elucidated.  Under the
    two-phase scenario, a significant fraction of mass growth during the second phase
    should be dominated by mergers with other galaxies that have low star formation.
    Hence, dry mergers, both minor and major ones, are 
    
    \begin{figure}[H]
    \centering 
    \includegraphics[width=8.75cm]{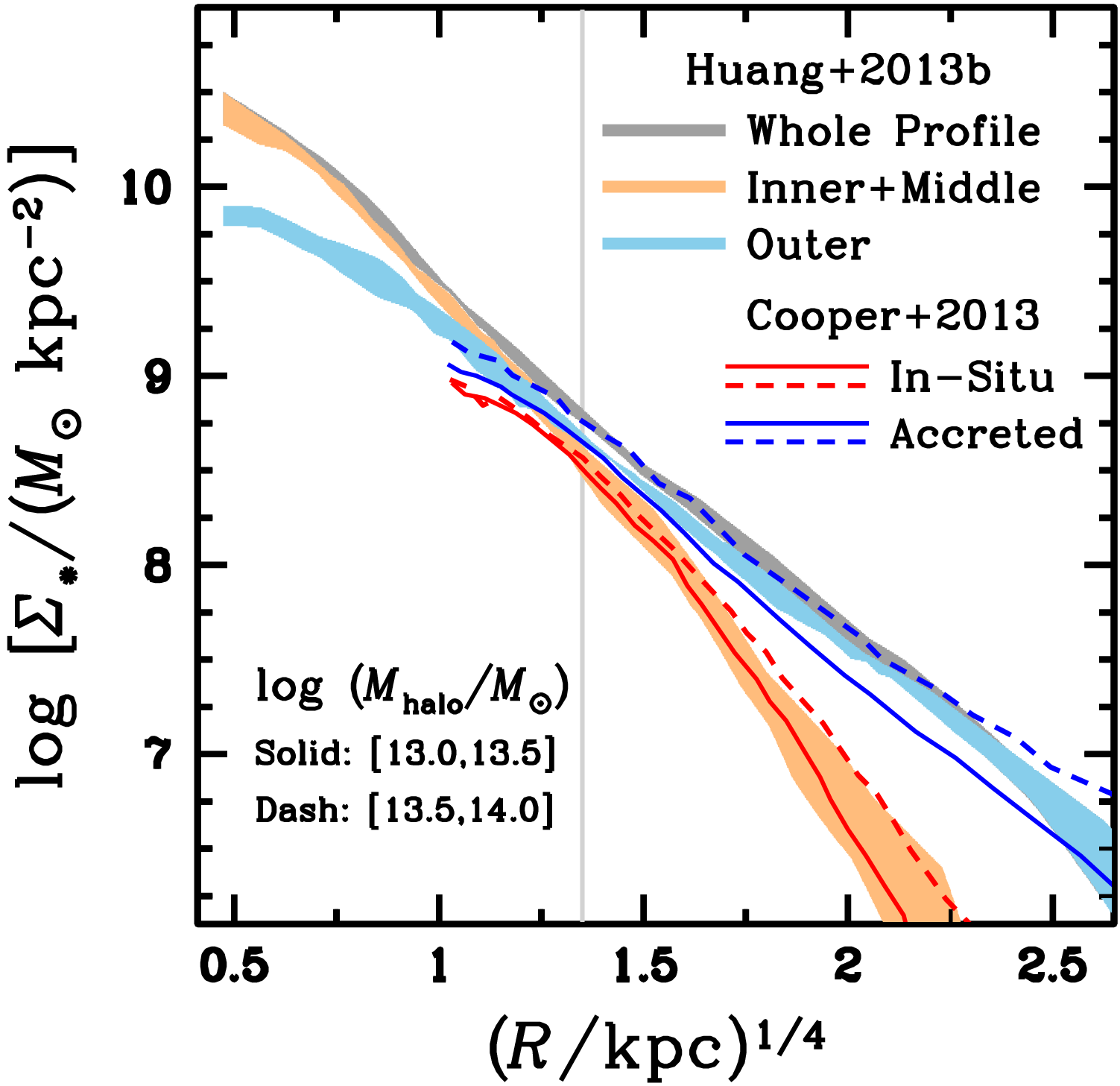}
    \caption{Comparison between the median stellar mass density profiles of the inner
        (orange shaded region) and outer (blue shaded region) photometric components of
        nearby massive ellipticals (Huang\etal 2013b), and the median profiles of the
        in-situ (red lines) and accreted  (blue lines) stellar components of model massive
        galaxies (Cooper\etal 2013).  Two halo mass bins are selected: $13.0 \le \log
        (M_{\rm halo}/M_\odot) \lt 13.5$ (solid lines) and $13.5 \le \log (M_{\rm
        halo}/M_\odot) \lt 14.0$ (dashed lines).  Profiles from simulation are more
        reliable at radii larger than the scale indicated by the vertical grey line.  We
        account for the difference in the choice of IMF used to estimate stellar mass.
        The profiles from Cooper\etal (2013) were derived assuming a Chabrier (2001) IMF,
        while the ones from Huang\etal (2013b) chose the default ``diet-Salpeter'' IMF
        from Bell\etal (2001); for a rough comparison, a consistent offset of $+0.2$ dex
        was added to the former.}
    \label{figure:2}
    \end{figure}
    
    \noindent crucial to the evolution of massive galaxies on the red sequence (e.g.,
    Bernardi\etal 2011). 
  
    Recent works by Huang\etal (2013a, b) clearly demonstrated that nearby massive
    ellipticals can be generally decomposed into two parts: a smaller, denser core (inner
    component) plus an extended stellar envelope (outer component).  The properties of the
    components indicate that they are in line with predictions of the two-phase scenario.
    If, indeed, more recent merger events were dominated by minor mergers and the stellar
    contents of these smaller systems are mainly redistributed out to large radii (e.g.,
    Hilz\etal 2012, 2013), then the color and luminosity of the outer component, properly
    separated from the the core, can provide an estimate on the average mass and number of
    minor mergers via the $M_{\ast}$-color relation, as illustrated by the cartoon in
    Figure~1.  Here, we demonstrate the feasibility of this novel approach by extending
    our 2-D image decomposition technique of Huang\etal (2013a) to multiple filters, and
    show that this is a promising tool for studying the merger history of ETGs. 

    The paper is organized as follows. Section 2 gives a brief overview of the sample of
    ellipticals and the photometric data.  We will also briefly describe the 2-D image
    decomposition method and the main results from $V$-band models.  Section~3 provides
    details about the methods we used to extend current models to other filters and
    describes the estimation of the $M_{\ast}$-color relation for nearby ETGs.  The main
    results are summarized in Section~4.  Section~5 briefly discusses the assumptions
    adopted in this work and possible future applications, ending with a summary in
    Section~6.
  
    Within this work, we assume $H_0$ = 70~km~s$^{-1}$ Mpc$^{-1}$, ${\Omega}_m=0.27$, and
    ${\Omega}_{\Lambda}=0.73$.
    
\section{Photometric Data of Nearby Massive Ellipticals}

    This work uses the multi-band optical images of nearby ellipticals from the
    Carnegie-Irvine Galaxy Survey (CGS; Ho et al. 2011, Paper~I).  CGS is a photometric
    survey of 605 bright ($B_T \lt 12.9$ mag), nearby (median $D_L$ = 24.9 Mpc) galaxies
    in the southern sky ($\delta \lt 0$\deg).  It was designed to provide an optical
    baseline sample to study in detail the structure of nearby galaxies with different
    morphologies.  The observations were made using the 100-inch du~Pont telescope at Las
    Campanas Observatory to provide high-quality \emph{BVRI} images.  The images have a
    size of 8\farcm9$\times$8\farcm9 and a pixel scale of $0\farcs259$.  Seeing conditions
    better than $\sim$1\asec\ were achieved for most of the observations.  The data were
    carefully reduced in a standard manner, as described in Paper~I and Li\etal (2011,
    Paper~II).  

    The morphology information for CGS galaxies was initially extracted from the Third
    Reference Catalogue of Bright Galaxies (de~Vaucouleurs\etal 1991), and later visually
    examined by the authors.  In general, $\sim$100 ellipticals are included in CGS\@.
    We managed to perform reliable image decomposition for 94 of them using their $V$-band
    images (Huang et al. 2013a, Paper~III).  In this work, we fully take advantage of
    these existing multi-component models to explore our proposed method for estimating
    the average merger mass ratio of nearby elliptical galaxies.   

    As a survey of very nearby galaxies using a relatively small CCD camera, the CGS data
    cannot compete with large imaging surveys (e.g., SDSS) in terms of sample size and
    accuracy in photometric calibration.  However, the greater image depth, higher spatial
    resolution, and better seeing conditions provided by CGS are crucial for reliable
    photometric modeling.  Due to the size of the CCD and the extended nature of the
    ellipticals, an accurate estimate of the sky background is nontrivial (in fact,
    surveys like the SDSS also suffer from over-subtraction of sky around bright
    ellipticals; see Blanton\etal 2011).  Meanwhile, slightly less than half of the CGS
    ellipticals were observed under non-ideal photometric conditions.  The calibration of
    these images is less accurate than others, which leads to slightly larger photometric
    errors.  However, we will show later that these data are sufficient to demonstrate the
    potential of our proposed method.  

    In Paper~III, we performed careful 2-D multi-component image decomposition on the
    $V$-band images of 94 CGS ellipticals, using the code {\tt GALFIT} (Peng\etal 2002,
    2010).  Contrary to common perception, we found that the 2-D surface brightness
    distribution of nearby ellipticals {\it cannot}\/ be described by a single \ser (1968)
    function.  All the sources require a multi-component photometric model.  In
    particular, the photometric models of 70 out of 94 ellipticals have three \ser
    components: (1) a compact central component with effective radius $R_e$ \lax\ 1 kpc
    comprising a small fraction of the light ($f \approx 0.1-0.15$); (2) an
    intermediate-scale component with $R_e \approx 2.5$ kpc and $f \approx 0.2-0.25$; and
    (3) an extended envelope with $R_e \approx 10$ kpc and $f \approx 0.6$, which
    is also moderately more flattened than the interior parts.

    These models were constructed in a purely empirical manner with no assumed physical
    meaning assigned to any component.  We started from a simple single-\ser model, and
    gradually built up the complexity by introducing more \ser components with reasonable
    constraints, until certain standards were reached. In Paper~III, we discussed at
    length the idea of a ``good photometric model.''  We concluded that, even for
    ellipticals, a model that follows the detailed 2-D surface brightness distribution is
    necessary to provide a complete picture of their structures.  More than that, we also
    demonstrated that the locations of the individual sub-components on the photometric
    projections of the fundamental plane are not only reasonable, but also physically
    meaningful.  The relevant procedures of model building will be described in the next
    section.  Paper~III provides more details of our image decomposition procedure. 

    In Huang\etal (2013b), we further compared these three-\ser models of nearby massive
    ellipticals with observations of high-$z$ compact, quiescent\footnote{In this work,
    ``quiescent'' means that the galaxy has very low star formation rate and does not
    contain a strong active nucleus.} ETGs.  The results strongly suggested that a
    hypothetical structure formed by the combination of the central and intermediate
    components (referred to collectively as the ``inner'' component in the rest of the
    paper) resembles the ETGs at $z \approx 1.5$ in many aspects, including the
    distribution on the $M_{\ast}$-size plane and the average mass density profiles
    outside the central kpc.  This evidence supports the idea that the inner part of
    nearby massive ellipticals can be regarded as the evolutionary descendants of the
    compact, quiescent galaxies formed at higher redshift under highly dissipative
    conditions.  On the other hand, the extended envelope we isolated from nearby
    ellipticals (referred to as the ``outer'' component here) follows a different
    $M_{\ast}$-size relation with larger scatter.  Compared to the inner component, we
    also noticed that there is a much more significant correlation between the size of the
    outer component and the total stellar mass (see Fig. 3 of Huang\etal 2013b), which is
    predicted by the two-phase scenario (Oser\etal 2010).  These pieces of evidence all
    indicate that the outer component can be seen as the by-product of non-dissipative,
    minor mergers.  

    Although the anatomy of nearby massive ellipticals is consistent with the expectations
    of the two-phase evolution model, it is still difficult to directly compare
    observations with simulations due to limitations on both sides.  In Cooper\etal
    (2013), the particle-tagging technique was combined with semi-analytic galaxy
    formation models (Guo\etal 2011) to predict the $M_{\ast}$ density profiles of massive
    galaxies in the Millennium~II simulation (Boylan-Kolchin\etal 2009).  The authors
    separate the stellar component formed through in-situ star formation from material
    accreted through mergers.  Figure~2 compares the median $M_{\ast}$ density profiles of
    the inner and outer components of nearby massive ellipticals ($M_{\ast} \geq
    10^{11.15} M_{\odot}$) from Huang et al. (2013b) with the average profiles of the
    in-situ and accreted components of simulated galaxies, in two relevant halo mass bins,
    from Cooper\etal (2013).  The similarity is striking in regions resolved in the
    simulations.  It is still too early to think that all observations can be perfectly
    explained, but we are motivated to apply our 2-D models to extract more physical
    insights. 
  
\begin{figure*}[tb]
    \centering 
    \includegraphics[width=16.5cm]{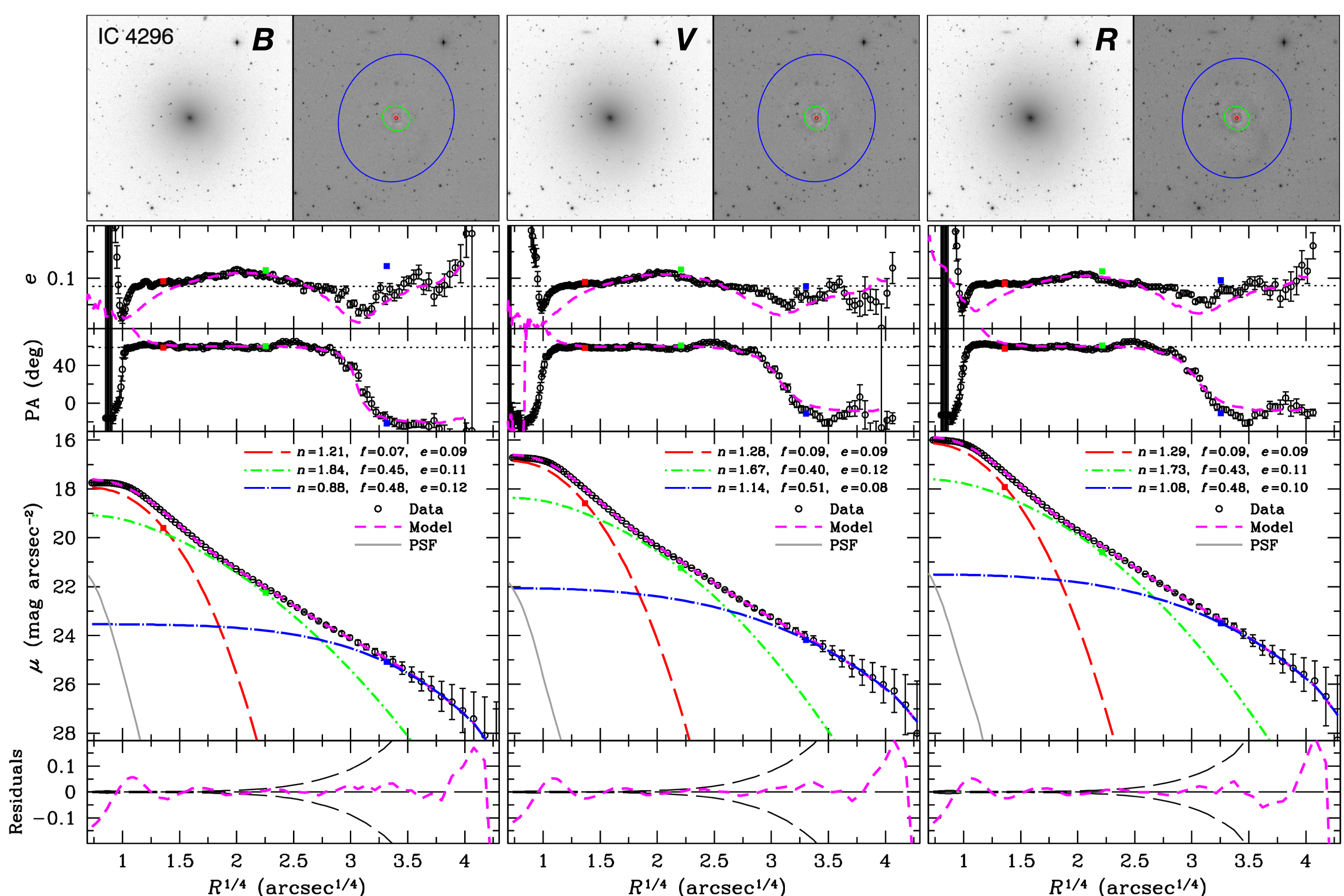}
    \caption{Comparison of the photometric decomposition of independently constructed
        three-\ser models for IC~4296 in the $B$ (left), $V$ (middle), and $R$ 
        (right) band.  The original image of the galaxy and the residual image (after
        subtracting the best-fit 2-D model) are displayed in the upper panels.  For the
        1-D information, the profiles for ellipticity, position angle, surface brightness,
        and residuals are shown from top to bottom.  The observed profiles along with
        their uncertainties are shown in black data points, while the profiles for the
        integrated models are shown as a pink dashed line.  The surface brightness profile
        for each \ser component is displayed in a different color; the same color scheme
        is used to trace the half-light radius in the residual images in the upper panels.
        The surface brightness profile of the PSF model in each filter is also shown (with
        arbitrary flux zero point).  For each component, we list the best-fit \ser 
        index $n$, light fraction $f$, and ellipticity $e$.}
    \label{figure:3}
    \end{figure*}

    \figurenum{3}
  
\section{Multi-component Decomposition in $B$ and $R$ Bands}

\subsection{Basic Ideas}

    In the ideal situation, multi-band images of the same galaxy can be modeled
    simultaneously under certain constraints.  Recently, this approach was realized
    through the efforts from the {\tt MegaMorph} team (H{\"a}u{\ss}ler \etal 2013;
    Vika\etal 2013, 2014) and has been applied to a large sample of galaxies (e.g.,
    Vulcani\etal 2014).  A similar approach has also been used to separate active galactic
    nuclei from their hosts (e.g., Bennert\etal 2011).  However, these methods so far only
    provide relatively simple modeling options.  Moreover, these techniques are still very
    time-consuming to apply to large images (in the case of CGS, 2048$\times$2048 pixels).
    In this work, we developed a very straightforward approach that provides suitable
    results for our purpose.  First, we confirm that consistent three-\ser models can be 
    independently obtained in all three filters (\emph{BVR} for most of the galaxies) then, 
    we use the three-\ser models of a reference filter as blueprint and apply it to the 
    other two bands under reasonable constraints.  

    Recent works (e.g., Vulcani\etal 2014) that fit single-\ser model profiles to ETGs
    find that their sizes systematically decrease with wavelength, indicative of a
    negative color gradient.  The number of major and minor mergers experienced throughout
    the lifetime of a galaxy spans a continuum, with ETGs occupying one extreme and disk
    and bulgeless galaxies lying on the other. In late-type galaxies, the reason that
    bulges can be differentiated from disks in their light profile and stellar content is
    that bulges and disks formed along different paths. If the negative color gradient of
    ETGs is the result of there being different stellar population components, then a
    multi-component analysis, similar to bulge-to-disk decomposition, might also bear that
    out. In Huang\etal (2013b), we showed that the inner component has properties that
    indeed closely match those of red nuggets. If this approach is sound, ideally we
    expect individual light profile shapes not to change across different filters, only
    the profile amplitudes. In reality the profile shapes do change due to dust or more
    complex mix of stellar populations, and this introduces uncertainty into the
    interpretation. 
  
    All ellipticals in Paper~III require multiple components, and the majority are best
    fit by a three-\ser photometric model.  For the 71 galaxies with reliable $V$-band
    decomposition, we start by independently building three-\ser models for them in the
    $B$ and $R$ bands, using the method described in Paper~III\@.  In our previous work, the
    $V$-band was selected mainly to minimize the ``red halo'' effect (e.g., Michard 2002;
    Wu\etal 2005) of the point-spread function (PSF), and also to take advantage of its relatively
    lower sky background.  To decide the average mass ratio of the 
    mergers using the mass-color relation, we choose to use the colors $B-V$ and $B-R$
    over $V-I$, as inclusion of a bluer filter makes them slightly more sensitive 
    to changes in the underlying stellar population (Bell \& de~Jong 2001).
    Hence, the mass-color relations using these two colors have steeper slopes than the one using
    $V-I$.  For the same uncertainty in the average color difference, this leads 
    to slightly smaller uncertainty in the derived mass ratio. As shown in Figure~A1
    of Paper~III, the red halo effect for the PSF is very small in $B$ and $R$.  It will
    still contaminate the color of the low-surface brightness region when its scale is
    similar to the size of the red halo, but for the ellipticals in our sample, the
    strongest part of the red halo (10\asec\ to 15\asec) will not fall on the low-surface
    brightness region.  Moreover, with the help of 2-D modeling, we no longer only look at
    the color at very large radii but the average color of the entire component.

    Compared to other systematics we face (e.g., photometric calibration, background
    subtraction), the bias introduced by the red halo effect can be ignored for these
    three filters (the same cannot be said for the $I$ band, which is why we do not use it
    in this analysis).  If there is still any, it should make the outer component slightly
    redder.  As discussed later, this will not affect our main conclusions.  

\subsection{Sky Background Estimation}

    Unlike the red halo effect, sky background subtraction is always a difficult issue to
    treat for massive ellipticals.  First and foremost, the extended outer envelope
    conspires with the limited field-of-view to make estimating accurate sky background
    difficult.  The difference in average sky level and spatial gradient of background
    among different filters further makes the background subtraction a major source of
    uncertainty for our models.  Under these circumstances, the background level and its
    gradient (as the field-of-view of our images is not very large, the background can be
    approximated as a tilted plane; fluctuations of higher order are ignored) become part
    of the model (see Yoon\etal 2011).  In principle, one can fully take this into account
    by sampling the whole parameter space using a Markov Chain Monte Carlo (MCMC) method
    (e.g., Yoon\etal 2011; D'Souza\etal 2014).  However, such approach is still extremely
    computationally expensive for large images and multi-component models
    ($\sim 20$ free parameters).  In this work, we compute a model-dependent
    background, in two ways.
  
    \begin{enumerate}
      \item First, we recalibrate the background of the $B$ and $R$ images using the same
          method described in the Appendix~B of Paper~III\@.  In short, we fit the image
          with a series of models with different number of components and a free
          background.  The average output background values from reasonable models and its
          scatter are taken as the best guess of the sky level and its uncertainty.
          During the formal three-\ser model fitting, we add a sky component with average
          intensity fixed at the best guess and free gradient in both the X and Y
          directions.  In practice, the spatial gradients of background are very small for
          our objects.    
     \item Even after the sky background has been re-calibrated, when we apply a
         three-\ser model from one filter as reference to another filter, we allow the sky
         value for the new filter to be either fixed or free.  For each choice of model,
         we make two estimates of color based on the stability of the background
         subtraction.   
    \end{enumerate} 

    Even after all these attempts, we still have two sources of uncertainties to worry
    about.  Sky subtraction is still the largest source of uncertainty for our three-\ser
    models.  As it significantly affects the surface brightness profiles at large radii,
    it mainly makes the magnitude and size of the outer component uncertain.  Therefore,
    the accuracy of sky subtraction in both bands basically controls the accuracy of the
    color difference between the inner and outer components.  At the same time, the
    uncertainty of photometric calibration applies to the image globally.  Hence, it is
    the main source of uncertainty for the absolute value of colors measured in this work,
    and determines how well we can locate a galaxy on the red sequence.  Despite these
    limitations, it will not stop us from demonstrating the proposed method.  As we will
    show later, the main results are not 

\begin{figure}[H]
    \centering 
    \includegraphics[width=8.35cm]{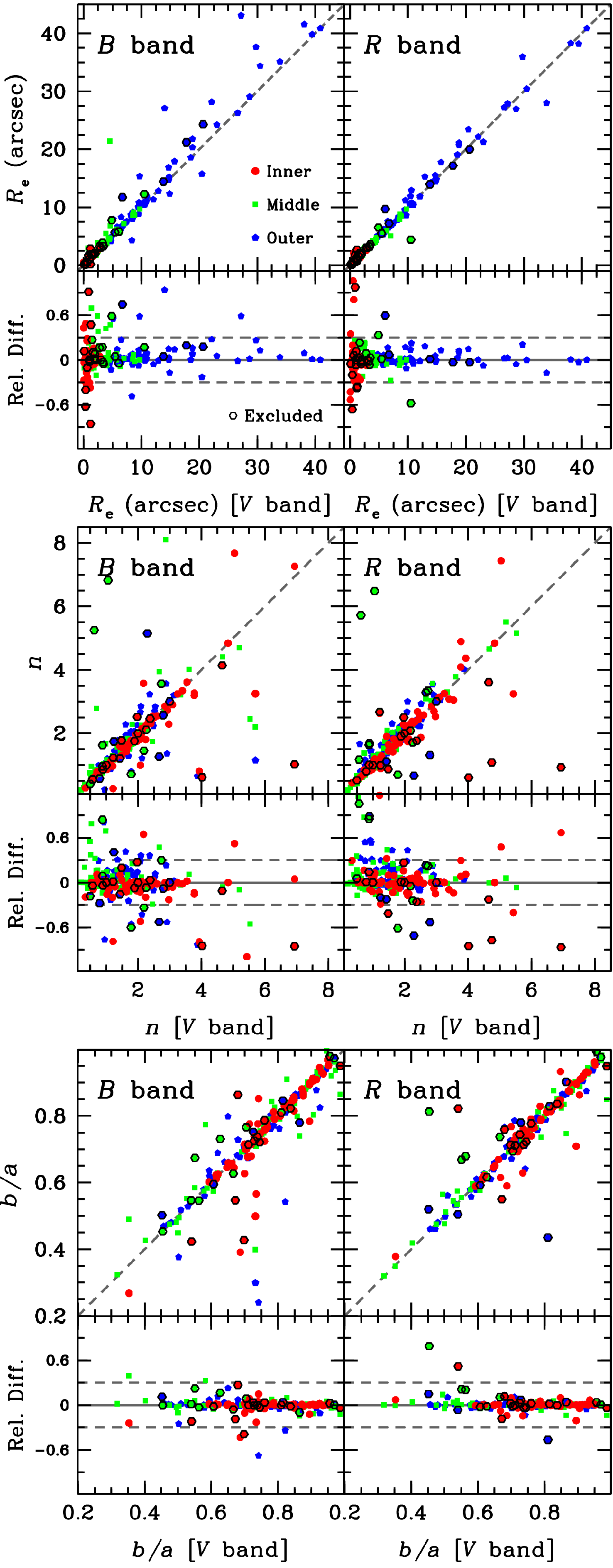}
    \caption{Comparison of the effective radius (top), \ser index (middle), and axis ratio
        (bottom) of the three photometric sub-components from the independent $B$-, $V$-,
        and $R$-band models.  The inner, middle, and outer components are displayed using
        red, green, and blue color.  The values for the $V$-band models are used as
        reference (X axis).  A black dashed line indicates the $x=y$ relation in each
        upper panel.  The differences of parameters derived from models in the $B$ and $R$
        band and the $V$-band reference models are displayed in the bottom panels.  Models
        that are excluded after this examination are highlighted using black outlines
        around data points for their components.    
    }
    \label{figure:4}
    \end{figure}

    \figurenum{4}
    
\begin{figure}[H]
    \centering 
    \includegraphics[width=8.65cm]{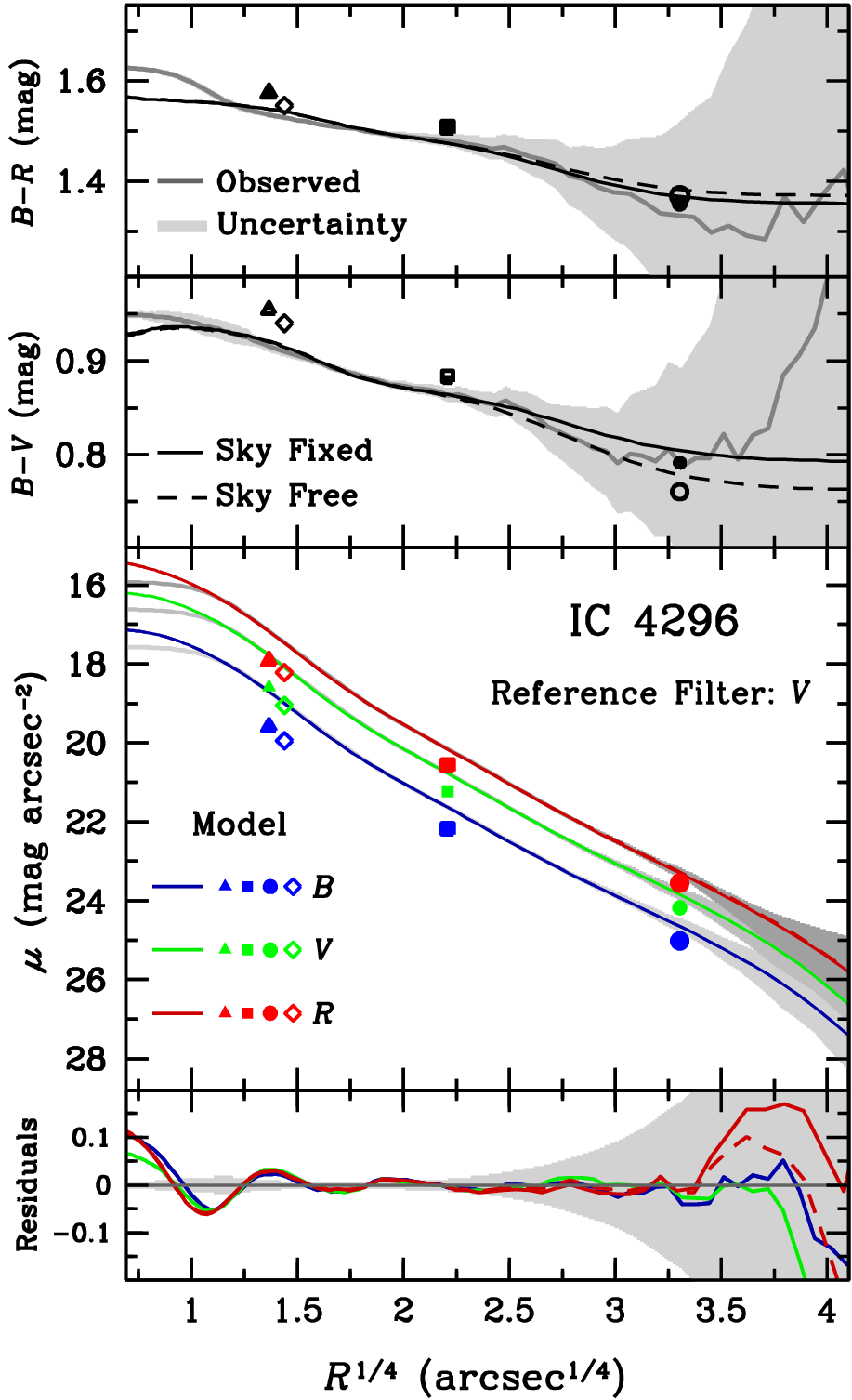}
    \caption{1-D representation of three-component \ser models for IC~4296 in the $B$,
        $V$, and $R$ bands.  From top to bottom, we show the 1-D $B-R$ and $B-V$ color
        profiles, the surface brightness profiles, and their residual profiles.  The
        observed surface brightness profiles and their uncertainties are shown in slightly
        different grey shaded regions.  The surface brightness profiles for models in $B$,
        $V$, and $R$ bands and their residual profiles are shown in different colors.  To
        highlight the properties of all three \ser component, different symbols are used
        to mark the effective radius and effective surface brightness of each component
        (inner: red; middle: green; outer: blue). The effective radius and the effective
        surface brightness of the composite inner$+$outer components are also highlighted
        using empty diamonds. In the two top panels, the observed 1-D color profiles
        without correction for PSF convolution are shown in grey solid lines; their
        uncertainties are shown using a shaded region.  The large uncertainties at large
        radii are partly due to the inclusion of uncertainties in sky background
        estimation.  The color profiles of the models are shown in black lines after
        corrected for PSF convolution.  Two different models are displayed here: fixed sky
        background (solid line) and free background (dashed line).  Similar to the surface
        brightness plot, the color of each component in the models are highlighted at its
        effective radius.  Filled symbols are used for fixed sky, and open ones are for
        free sky.  
    }
    \label{figure:5}
    \end{figure}

    \figurenum{5}
    
    \noindent affected.  In the future, this situation will be improved by using images
    with larger sky coverage and more advanced methods in image modeling. 

\subsection{Procedures for Model Fitting}

    Using the updated sky background value, we first try to independently build a
    three-\ser model in the $B$ and $R$ bands.  The PSF model and object mask are prepared
    exactly as for the $V$-band images in Paper~III\@.  To make these models more
    objective, we do not start from best-fit parameters from the $V$ band.  Instead,
    initial guesses for model parameters are also decided in the way described in
    Paper~III\@.  In general, the results are very encouraging.  For most ellipticals with
    $V$-band three-\ser models, similar models can be obtained in both the $B$ and $R$
    bands.  Figure~3 shows an example for IC~4296.  From the upper panel, we can see that
    the residuals are very good in all three filters.  And judged by the surface
    brightness profiles and key parameters for the three \ser components, the models in
    all three bands are indeed very consistent.  In Figure~4, we further compare the
    effective radius ($R_{e}$), \ser index ($n$), and axis ratio ($b/a$) for each
    component in models from the three filters, and show the relative difference using the
    value from $V$-band as reference.  For most galaxies, we have very consistent models
    in all three bands.  In general, the $R$-band models are more similar to the $V$ band
    than the $B$ band, especially for $R_{e}$ and $b/a$.  This is expected, for three
    reasons: (1) The typical signal-to-noise ratio at large radii and the typical
    background level is quite different between $B$ and the two redder bands, which can
    affect $R_{e}$ and $n$ for the outer component.  (2) We identified dust features
    around the center of 21 Es in our sample.  However, we will not be able to resolve
    dust features at smaller scales, as those detected by high-resolution {\it HST}\/
    images (e.g., Lauer\etal 2005).  Dust naturally has a greater effect at shorter
    wavelengths.  (3) The typical $B$-band PSF is systematically broader than that in the
    $V$ and $R$ bands.  Of course, it is also possible that the differences in $R_{e}$ and
    $n$ simply reflect some intrinsic wavelength dependence for individual components.  We
    cannot resolve this given the current uncertainties.   

    After examining these models, we exclude a few galaxies with large differences between
    their $V$- and $R$-band results, which may be due to problematic background estimation
    in either of the filters; $B$-band models are not considered in this process. We
    mostly focus on the intermediate and outer components during the examination.  The
    large uncertainties for parameters of the central component can be easily affected by
    dust features, unresolved structure, or tiny PSF errors.  As discussed in Paper~III,
    even though the central component tends to have large relative uncertainties (in some
    cases, the central components are not well-resolved by our observations), they are
    very much needed to achieve a satisfactory global model.  The final sample contains 60
    objects.

    We use the best-fit three-\ser $V$-band models from Paper~III as reference to generate
    corresponding models in $B$ and $R$.  In light of the main purpose of this work, all
    parameters for the three \ser components are fixed except for their total magnitude
    (basically a color term is added to each component).  As mentioned, a sky component is
    included, and its value is either fixed to the updated value or set free; sky
    gradients in both X- and Y-direction are left free for both cases.  The background
    gradients are always quite small for these galaxies.  Models without sky gradient
    essentially show no difference.  After the models are built, the color for each
    component is extracted and the overall color profile is generated; the PSF convolution
    in both filters is removed before the color profiles are made.  In Figure~5, the
    one-dimensional (1-D) surface brightness profiles of models in all three filters,
    their residual profiles, and the $B-V$ and $B-R$ color profiles are displayed along
    with the $R_{e}$, the surface brightness at $R_{e}$, and the color of each components.
    The residual profiles in the three bands are very good and consistent.  More
    importantly, both the $B-V$ and $B-R$ color profile of our three-\ser component models
    closely follow the 
    
\begin{figure*}[bt]
    \centering 
    \includegraphics[width=16.5cm]{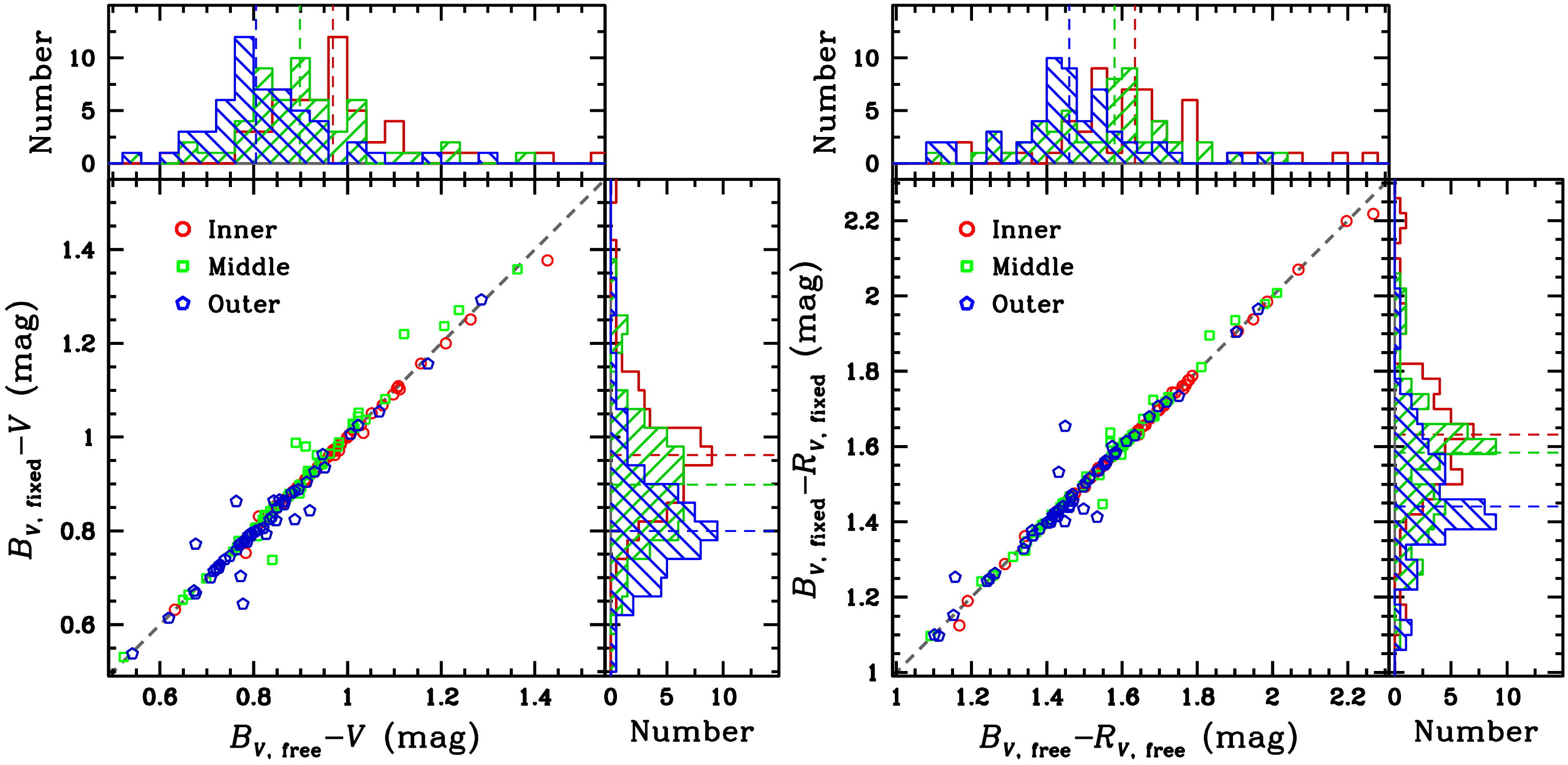}
    \caption{Distributions and comparisons of $B-V$ (left) and $B-R$ (right) colors of the
        inner (red circle), middle (green square), and outer (blue pentagon) components
        from the sky-free and sky-fixed models.  A grey dashed line is shown to indicate
        the $x=y$ relation.  For the histogram, the median color of each component is also
        highlighted using a dashed line with corresponding color.  
    }
    \label{figure:6}
    \end{figure*}

    \figurenum{6}
    
    \noindent observed one.  Differences in the innermost region ($\sim 1$\asec) are due
    to different treatment of PSF convolution.  
  
    To extract the observed color profile from the data for comparison with the model
    fits, one has to account for the PSF differences between the different filters.  We
    choose a commonly adopted method: convolve the image in one filter with the PSF model
    from the other filter before the color profile is extracted.  The distinct bump of the
    $B-R$ profile in the center is likely an artefact, and it has little effect on the
    color profile of the model.  On the other hand, at very large radii, usually close to
    the edge of the image, the observed $B-V$ color profile sometimes shows a weird upward
    ``U-shape'' feature, which is unlikely to be real.  As discussed in Section~3.1, it is
    also not caused by the PSF red halo effect, considering its large radius, but is
    probably due to slight over-subtraction of background in the $V$ band.  Due to the
    large uncertainty of the background in $V$, the color of the outer component also
    clearly varies with fixed or free background.  In fact, this is an inherent weakness
    of the 1-D color profile analysis, as it is extremely sensitive to not only the
    structure of the PSF but also the accuracy of background subtraction.  Meanwhile, the
    color of the outer component from our PSF-convolved 2-D model is much more stable
    against these issues.
    
    In Figure~6, the $B-V$ and $B-R$ color from the fixed and free sky modeling are
    compared for each component.  The overall agreement is good.  Large deviations are
    mostly seen for the outer component alone, and error in sky determination for at least
    one of the filters is likely to be the main reason.  We also relax the parameter
    constraints, where both the \ser index and total magnitude of each component are free
    during the fitting.  As expected, the extra degrees of freedom lead to minor
    improvement in the residual, mostly in the central region, which affects the \ser
    index the most. However, it leads to very little to no difference in the inferred
    average color profile.  We therefore stick to the multi-band models where only the
    magnitude of each component is allowed to vary between the filters.  The color
    information for the sample is summarized in Table~1.   
   
\begin{figure*}[bt]
    \centering 
    \includegraphics[width=16.5cm]{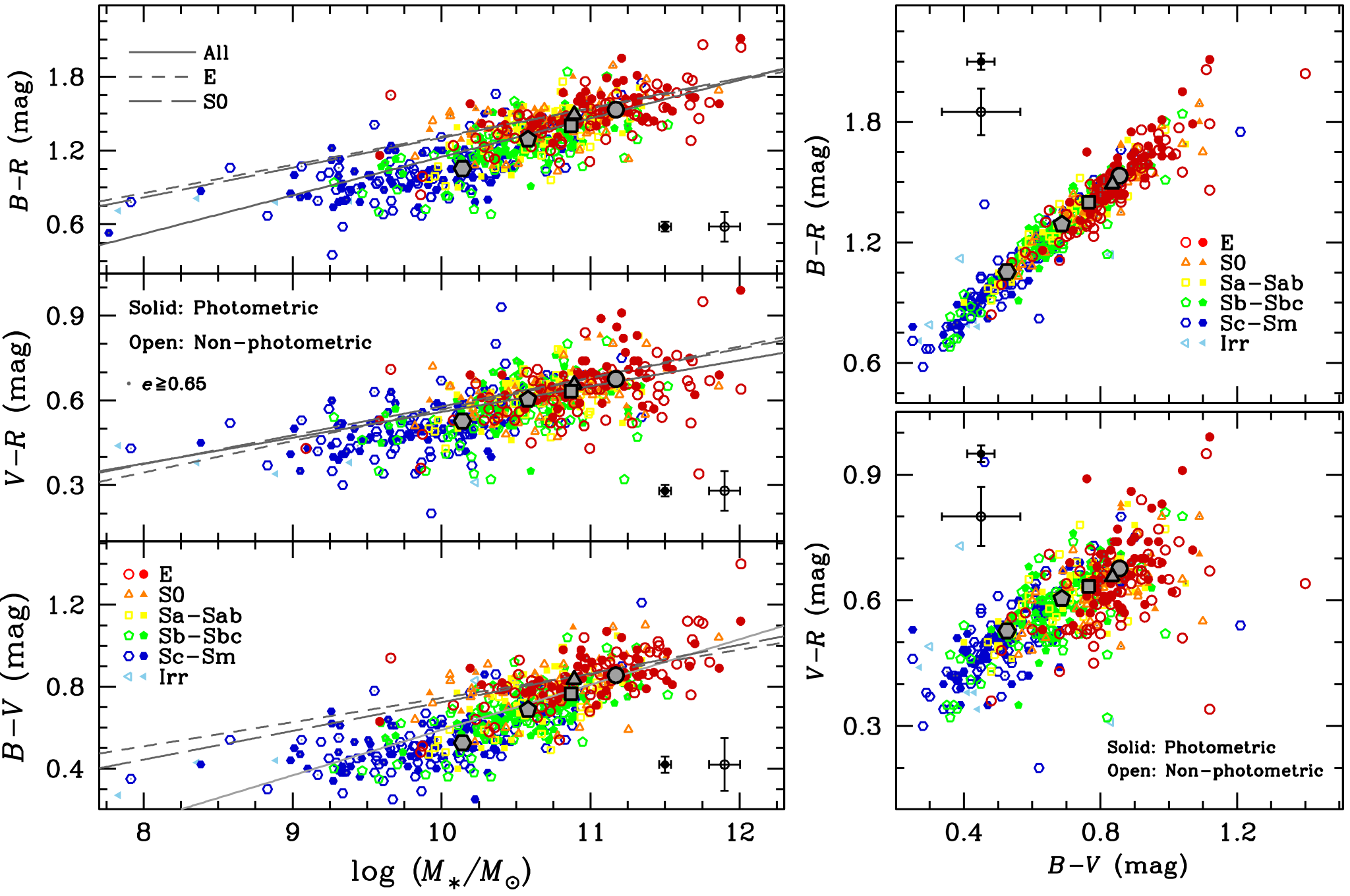}
    \caption{The $M_{\ast}$-color and color-color relations for all CGS galaxies.  On the
        left panel, from bottom to top, the $M_{\ast}$-color relations for $B-V$, $V-R$,
        and $B-R$ colors are displayed.  On the right side, the relations between $B-V$
        and $B-R$ (top) and $V-R$ (bottom) colors are shown.  Different colors and symbols
        are used for galaxies with different morphologies, as shown in the legends.
        Filled and open symbols are used to separate galaxies observed under photometric
        and non-photometric conditions, respectively.  The median stellar mass and optical
        color of each morphological type is shown using large grey symbol.  Typical errors
        for the photometric and non-photometric subsamples are shown on each plot.  On all
        $M_{\ast}$-color relations, three different lines highlight the general mass
        dependence for the ellipticals (dashed line), S0s (long-dashed line), and the
        whole sample (solid line) using simple linear fitting results. 
    }
    \label{figure:7}
    \end{figure*}

    \figurenum{7}

\begin{figure*}[bt]
    \centering 
    \includegraphics[width=15.5cm]{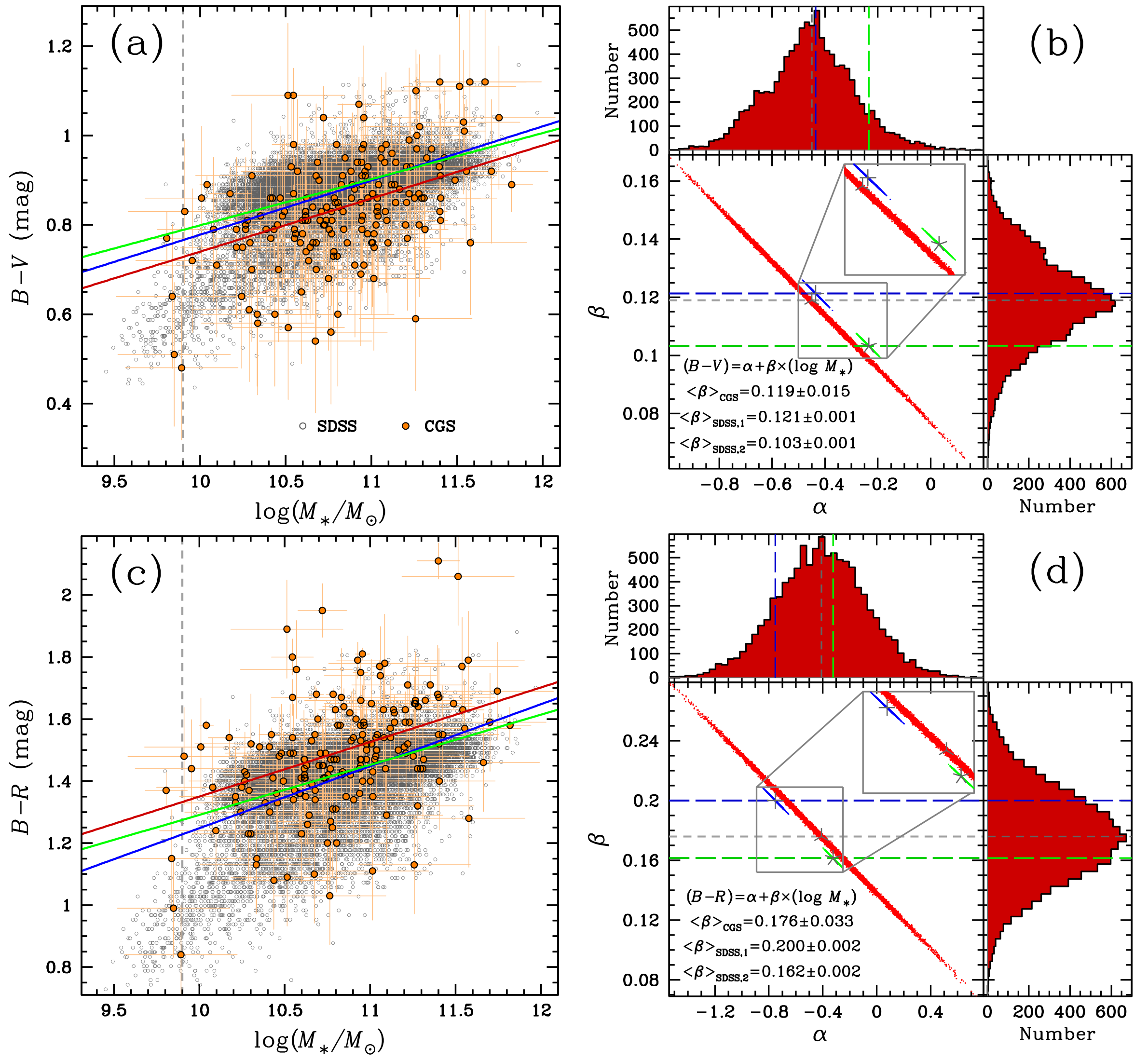}
    \caption{The $M_{\ast}$-color distributions of CGS ETGs and nearby SDSS ETGs, and
        their best-fit linear relations.  Panels (a) and (c) show the distributions of
        SDSS (grey, open circle) and CGS (orange, filled circle, with error bar) galaxies
        on the $M_{\ast}$-$(B-V)$ and $M_{\ast}$-$(B-R)$ planes.  A grey dashed line
        indicates the applied $M_{\ast}$ cut during the fitting.  Panels (b) and (d)
        display the distributions of slope ($\beta$) and intercept ($\alpha$) from the
        MCMC fitting of the relations for CGS (red), SDSS $p_{{\rm E}+{\rm S0}} \gt 0.7$
        (green), and SDSS $p_{{\rm E}+{\rm S0}} \gt 0.5$ (blue) samples.  The final
        estimation of the slopes are shown on the plots, and also highlighted using
        different lines.  Three solid lines with corresponding colors are displayed on
        panels (a) and (c) to illustrate the best-fit relations. The red histograms show
        the posterior distributions of parameters from a $N_{\rm MCMC}=10000$ MCMC run. 
    }
    \label{figure:8}
    \end{figure*}

    \figurenum{8}
   
\subsection{$M_{\ast}$-color Relations} 

    To interpret the color offset between the inner and outer components, we compare our
    measurements with the slope of the $M_{\ast}$-color relation followed by nearby ETGs.
    According to recent observations, massive quiescent galaxies span a sequence on the
    $L_{\ast}$-color or $M_{\ast}$-color plane, up to $z\approx1.5$ in both overdense and
    field-like environments (e.g., Tanaka\etal 2005; Nicol\etal 2011).  The slope of the
    $L_{\ast}$-color relation for these galaxies appears to have not changed much since $z
    \approx 1.0$.  The evolution of the intercept of the relation favors a quick
    exponential decline of star formation rate (e.g.,  Fritz\etal 2014).  This is in line
    with the two-phase formation scenario of massive galaxy evolution.  
  
    Figure~7 displays the distributions of all CGS galaxies, according to their morphology, 
    on different $M_{\ast}$-color and color-color planes.  Stellar mass for CGS galaxies
    is derived using the empirical relation between the $B-V$ color and stellar $M/L$ in
    Bell\etal (2003).   To compare with other samples, a $-0.2$ dex constant offset is
    applied to make our masses compatible with a Chabrier (2003) stellar initial mass
    function (IMF); this is a rough approximation, but it has no effect on the slope of
    $M_{\ast}$-color relation.  The photometric uncertainty of the CGS data leads to
    larger scatter, but the behavior of each morphology class is reasonable.  We will
    first fit the $M_{\ast}$-color relation using CGS data alone.  The ETG sample
    comprises all ellipticals and S0s.  Under the assumptions of the two-phase formation
    scenario, most of the mergers during the last phase should be dry (e.g., Tal\etal
    2009), and hence it is natural to consider only ETGs, systems that closely track the
    red sequence.

    We adopted the Least Trimmed Squares (LTS) algorithm used in Cappellari\etal (2013),
    as implemented in the {\tt IDL} version of the code {\tt lts\_linefit} from Michele
    Cappellari's website\footnote{http://www-astro.physics.ox.ac.uk/$\sim$mxc/software/};
    the code is very robust against (the possibly large number of) outliers and can
    properly account for uncertainties from both axes.  Meanwhile, {\tt lts\_linefit} also
    provides the option to use {\tt linmix\_err} by Kelly (2007) to derive the best-fit
    relation and its uncertainty.  We use 10,000 MCMC sampling to get the posterior
    distribution of parameters in {\tt linmix\_err}.  To reduce the influence from
    outliers, 3-$\sigma$ clipping is performed before the fitting.  We also impose a lower
    stellar mass limit of $10^{9.9} M_{\odot}$ to exclude a few low-mass galaxies with
    suspicious classification.  For CGS data, the errors on the colors and stellar mass
    are all dominated by the systematic uncertainties from photometric calibration.  These
    errors are clearly highly correlated due to our photometry-based approach to derive
    stellar mass.  The results are shown as part of Figure~8.  For the $B-V$ color, the
    best-fit result is

      \begin{equation}
          (B-V) = (0.119\pm0.015) \times\log M_{\ast} - (0.450\pm 0.160),
      \end{equation}

    \noindent and the corresponding relation for the $B-R$ color is

      \begin{equation}
          (B-R) = (0.176\pm0.030) \times\log M_{\ast} - (0.410\pm 0.330).
      \end{equation}

    \noindent As expected, the CGS data result in large uncertainties on the slopes and
    intercepts of the above relations.  Also, {\tt lts\_linefit} results without invoking
    {\tt linmix\_err} are consistent with above ones.

    To ensure that the derived slopes are useful, we seek help from the SDSS, as it
    provides a more complete, much larger sample of nearby galaxies with uniformly
    calibrated photometry.  S\'ersic-fitting photometry of galaxies within $0.05 \le z \le
    0.07$ was selected from the catalog of Simard\etal (2011).  This redshift range
    ensures that galaxies are nearby enough to be well resolved and can be compared with
    our CGS sample, while not so nearby in order that the high-mass population can be
    sampled.  We convert the extinction- and $K$-corrected $g-r$ color, derived from \ser
    fitting, to $B-V$ and $B-R$ colors using the empirical equations from
    Lupton\etal(2005).  

    This sample is further cross-matched with the spectrophotometric stellar mass
    estimates based on the Simard\etal models (Mendel\etal 2014).   We assign a constant
    0.15 dex uncertainty.  To obtain a sample of ETGs (ellipticals $+$ S0s), we use the
    automated morphological classifications from Huertas-Company\etal (2011), who apply a
    threshold of $p_{\rm E+S0} \gt 0.7$ to isolate likely ETGs.  This produces a sample of
    16,100 nearby ETGs, which is compared in Figure~8 along with the CGS sample.  The
    general distributions and $M_{\ast}$ dependence are very similar, though it is likely
    that a small constant shift in either mass or color direction can make the agreement
    better.  The CGS $B-V$ color is a bit ``too blue'' when compared with SDSS at similar
    $M_{\ast}$, while the CGS $B-R$ color is ``too red.''  This can be easily due to
    difference in filters and IMF choice.  The $M_{\ast}$-color (converted $B-V$ and
    $B-R$) relations of the SDSS sample are fit in exactly the same way.  The best-fit
    relation for $B-V$ color is

      \begin{equation}
          (B-V) = (0.103\pm0.001) \times\log M_{\ast} - (0.234 \pm 0.012),
      \end{equation}
  
    \noindent and for $B-R$ color 

      \begin{equation}
          (B-R) = (0.162\pm0.002) \times\log M_{\ast} - (0.324 \pm 0.020).
      \end{equation}

    \noindent As shown in Figure~8, the slopes from SDSS sample are slightly shallower,
    but can still be seen as consistent with CGS within the uncertainties.  The intercepts
    of the SDSS sample show systematic offsets that confirm the direction of overall
    shifts between the distributions of these two samples.  Although the overall agreement
    is quite encouraging, especially considering the systematic differences involved, we
    note that the slopes are quite sensitive to the choice of morphological threshold.  If
    we switch to $p_{\rm {E+S0}} \gt 0.5$, more S0s and other early-type disk galaxies are
    included, and the slopes for both relations become much steeper: $\beta = 0.122$ for
    $B-V$ and $\beta = 0.197$ for $B-R$.

    Along with the observed uncertainties, {\tt lts\_linefit} can also account for the
    intrinsic scatter of the relation, which is important for estimating the uncertainty
    of the mass ratio we are deriving.  However, for the relations based on CGS data, the
    intrinsic scatter is strongly affected by systematic uncertainties.  In the case of
    the SDSS data, the derived intrinsic scatters are always very small.  Although it is
    known that the red sequence is very tight at the high-mass end, we realize that the
    uncertainty of the slope of the $M_{\ast}$-color relation is dominated by the choice
    of which ``early-type'' galaxies we include in the sample.  When more S0 galaxies are
    included, the color distribution at the low-mass end shifts slightly bluer.  This
    leads to a steeper slope of the $M_{\ast}$-color relation, as we just showed.
    Although it is safe to assume that the late-time evolution of massive ellipticals is
    dominated by non-dissipative mergers, we still know very little about the detailed
    morphology of the smaller satellites involved in the merger.  This uncertainty is
    interesting, and affects the final estimation of the merger mass ratio.  We will
    return to this point later.         

\begin{figure*}[bt]
    \centering 
    \includegraphics[width=16.5cm]{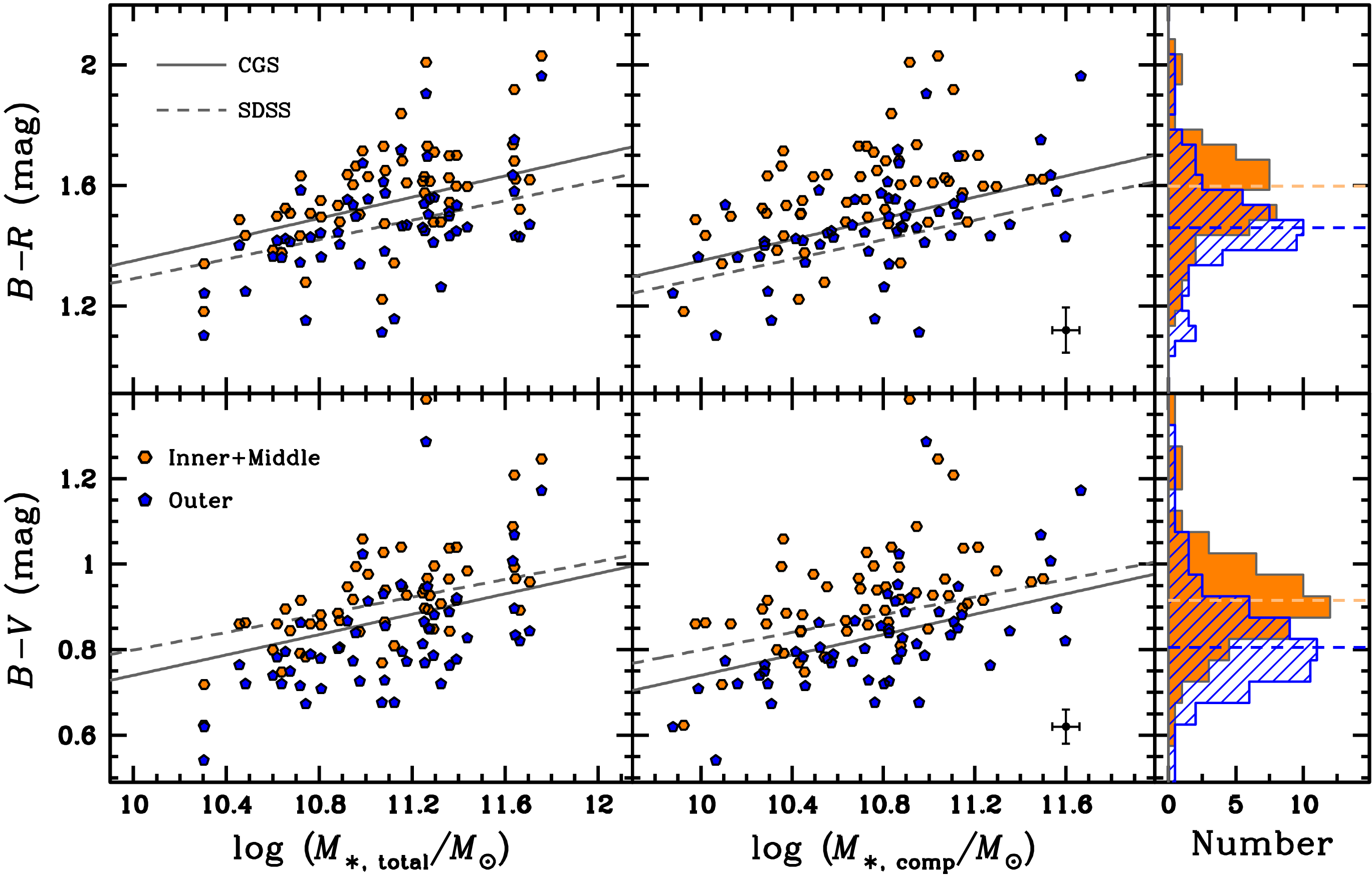}
    \caption{Distributions for colors ($B-R$: top; $B-V$: bottom) of the inner (orange)
        and outer (blue) components, along with their relations with the total stellar
        mass of the galaxies ($M_{{\ast}, {\rm total}}$; left) and the stellar mass of
        each component ($M_{{\ast}, {\rm comp}}$; middle).  The best-fit relations of CGS
        ETGs (solid line) and SDSS $p_{{\rm E}+{\rm S0}} \gt 0.7$ sample (dashed line) are
        displayed on the left and middle panels.  On the histograms on the right, dashed
        lines with corresponding colors are used to highlight the median value of the
        color distributions.  
   }
   \label{figure:9}
   \end{figure*}

    \figurenum{9}

\section{Results} 
  
    With the photometric models for the images in the three bands and the $M_{\ast}$-color
    relations in hand, the main  results of this work are very straightforward to obtain.
    In the following two sections, we first briefly summarize the color difference between
    the inner and outer components, and then estimate the average merger mass ratio based
    on our method. 

\subsection{Color Difference between Inner and Outer Components} 
  
    In Section~3, we showed that the well-known negative color gradient in ellipticals can
    be accurately recovered by a combination of three \ser components with very similar
    properties.  This suggests that, as expected, the color of the outer portion should be
    bluer than the portion at small radii.  Here we explore this in detail.  First, we
    compute the integrated color of the hypothetical inner component
    (central$+$intermediate) by simply combining their luminosities together.  The
    properties of this structure are known to be similar to those of high-$z$ compact,
    quiescent galaxies, which are likely to be the progenitors of the massive ellipticals
    in our sample;  they also remind us of the predicted in-situ component in the
    two-phase scenario for the formation of massive galaxies (Huang\etal (2013b and
    Figure~1).  Meanwhile, the outer component may well represent the accreted stellar
    envelope.  We first examine their color distributions and their behavior on the
    $M_{\ast}$-color plane (Figure~9). We consider both the total $M_{\ast}$ and the
    stellar mass for each component.  We estimate $M_{\ast}$ of the inner and outer
    components using their own colors, as derived from models with $V$ band as reference
    and free sky background in the other filter, in combination with the recipe for $B-V$
    color from Bell\etal (2003).  The total stellar mass $M_{\ast, {\rm total}}$ is the
    combination of the inner and outer components.  The $M_{\ast}$ estimated based on the
    $B-R$ color is systematically higher by $0.08\pm0.05$ dex; the difference does not
    depend on galaxy luminosity, and the choice of recipe here does not affect any of our
    results.  Recently, Roediger \& Courteau (2015) demonstrate that the $M_{\ast}$
    estimated based on color-$M/L$ relation typically has a random error of 0.10--0.14
    dex, and is not noticeably worse than the one from spectral energy distribution (SED)
    fitting, given accurate photometry.

    In the $M_{\ast, {\rm total}}$-color relations, the outer components are distributed,
    on average, slightly below the relation defined by CGS ETGs.   A few ellipticals on
    the high-mass end show suspiciously red colors.  Two of them (NGC~4760 and NGC~7145)
    were observed under non-photometric conditions, and hence their colors may be
    suspicious.  In any case, we will rely mostly on the statistical average properties,
    which should be more robust.  When the inner and outer components are placed on the
    $M_{\ast}$-color relations using their own stellar mass, they both seem to follow
    relations that share very similar slopes with the ones defined by the local ETG
    population.  Importantly, the outer components now almost perfectly follow the global
    $M_{\ast}$-color relations for ETGs.   This is consistent with our assumption that the
    outer part grows by merging with other galaxies that are {\emph{already}} on the red
    sequence.  

    The distributions of the color difference between the inner and outer components, and
    their dependence on $M_{{\ast}, \rm total}$, are shown in Figure~10.  For further
    analysis, we split the sample into high-mass and low-mass subsamples, using the median
    total stellar mass of $10^{11.10} M_{\odot}$.  Median and mean color differences are
    estimated for them separately (the green and orange points).  The high-mass subsample
    is most relevant for the two-phase formation scenario.  According to recent works
    (e.g., Patel\etal 2013), these very massive ellipticals are likely to be the direct
    descendants of high-$z$ red nuggets.  As the most massive systems, they are more
    likely to have experienced more accretion events.  Their photometric properties from
    our decomposition are also systematically different from those of lower mass objects,
    being more consistent with theoretical expectations for 
    
\begin{figure}[H]
    \centering 
    \includegraphics[width=8.75cm]{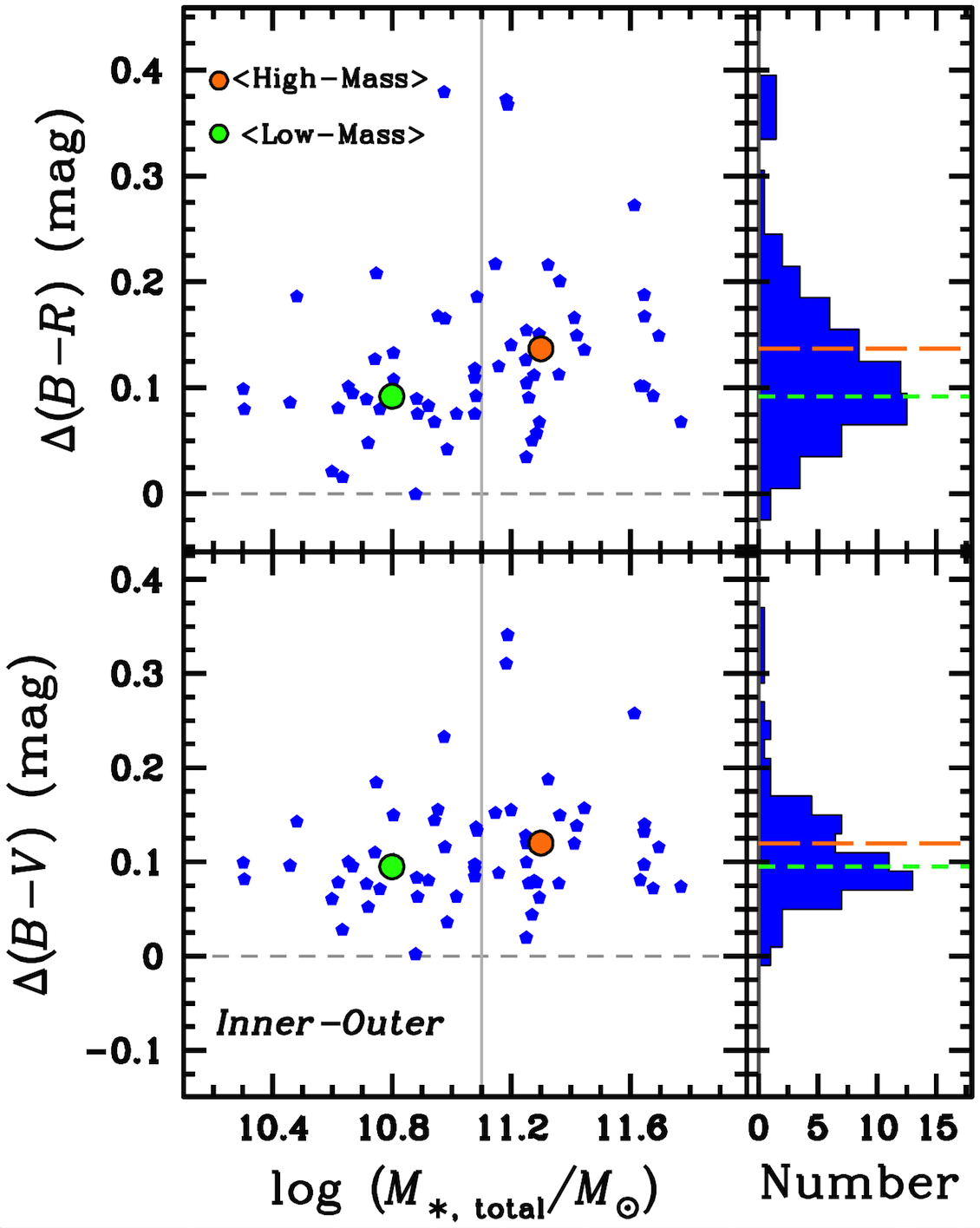}
    \caption{Distributions of color difference between inner and outer components and
        their relations with the total stellar mass.  The samples are separated into
        low-mass and high-mass subsamples using their median value of $M_{{\ast}, {\rm
        total}}$ (vertical grey solid line).  The median value of $M_{{\ast}, {\rm
        total}}$ and color difference for both subsamples are highlighted using green
        (low-mass) and orange (high-mass) filled circles.
    }
    \label{figure:10}
    \end{figure}

    \figurenum{10}
    
    \noindent the two-phase formation scenario (Paper~III and Huang\etal 2013b).  The
    choice of $M_{\ast}$ boundary is a bit arbitrary, but the mass range for the high-mass
    subsample turns out to be appropriate.  We chose the models with sky parameter left
    free because they contain fewer extreme outliers, but the sky-fixed models give
    consistent results.

    The median (average) color difference for the whole sample is $\Delta (B-V) = 0.10$
    (0.11) mag.  For the high-mass subsample, the difference is slightly higher, $\Delta
    (B-V) = 0.12$ (0.13) mag for the  median (average), whereas for the low-mass subsample
    $\Delta(B-V) = 0.10$ (0.10) mag.  The $B-R$ color behaves similarly.  The color
    difference for the high-mass subsample, $\Delta(B-R) = 0.14$ (0.14) mag is
    systematically larger than that for the low-mass subsample, $\Delta(B-R) = 0.09$
    (0.10) mag.  It is worth pointing out that the color difference between sub-components
    derived from 2-D models is not equivalent to the color difference within a radius
    range derived from the 1-D color profile.  The two sub-components highly overlap with
    each other, just as predicted for the in-situ and accreted components in simulations
    (e.g., Cooper\etal 2013;  Rodriquez-Gomez\etal 2015).  Therefore, the 1-D color
    difference across the overlapped region should be smaller than the color difference
    between the average color of the inner and outer components.  In Appendix~A, we
    compare the 1-D color profiles from our 2-D models with previous works, and show that
    the 1-D color difference based on our models are consistent with the median color
    profiles of galaxies in the same stellar mass range.  The average 1-D color difference
    between 3 and 30 kpc (or between 0.5 to 4 times $R_{\mathrm{e}}$) is indeed smaller
    than the $\Delta (B-V)$ and $\Delta (B-R)$ between inner and outer components.
    Assuming the in-situ and accreted components are reasonably represented by our models,
    this suggests that one has to go to radius larger than $\sim 30$ kpc or $\sim 4
    R_{\mathrm{e}}$ to find regions that are completely dominated by the accreted
    component; at the same time, the contribution of accreted stars to the projected
    stellar mass density inside $R_{\mathrm{e}}$ still cannot be completely ignored (e.g.
    Rodriquez-Gomez et al. 2015).

\begin{figure*}[bt]
    \centering 
    \includegraphics[width=16.5cm]{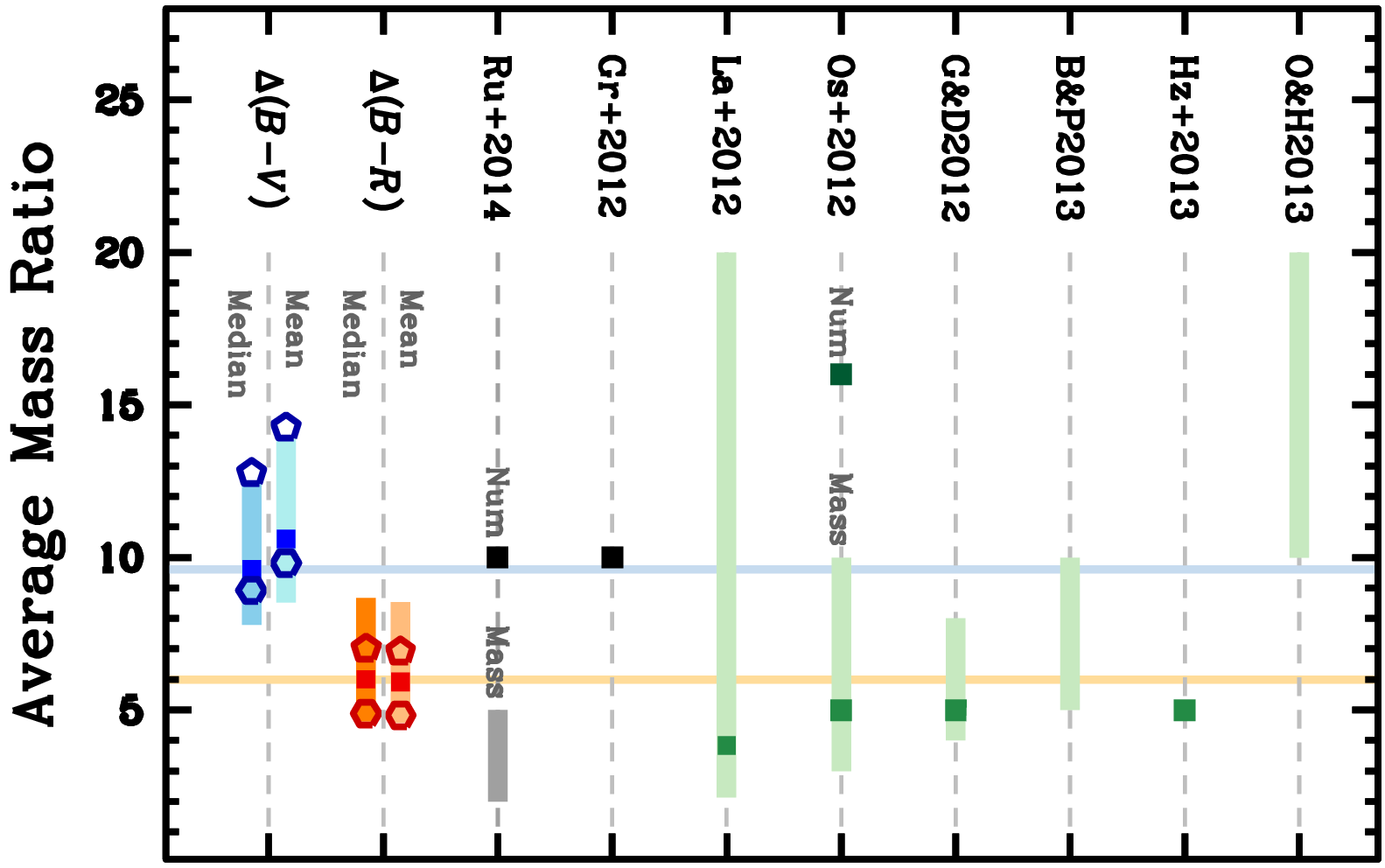}
    \caption{The distribution of average stellar mass ratio from this work and from various
       literature sources.   The estimations using $B-V$ (blue) and $B-R$ (red) colors
       from this work are displayed on the left side, using both the median and mean color
       difference.  Solid squares in darker colors are used for the estimations using the
       best-fit CGS relations; a lighter, shaded bar illustrates the allowed mass ratio
       from the 1$\sigma$ uncertainties of the relations.  Open pentagons and hexagons
       with corresponding colors are used for the estimations using the best-fit slopes
       for the SDSS $p_{{\rm E}+{\rm S0}} \gt 0.7$ and $\gt 0.5$ samples.  The estimations
       using median color difference and best-fit CGS relations are used as reference.
       Two horizontal solid lines with slightly lighter colors are displayed to compare
       with other estimations.  Similar estimations in recent observational works are
       highlighted using grey squares and vertical bars, while estimations from numerical
       simulations are shown in the same way using different green colors.  See the text
       for detailed explanations of each reference used here.  Reference code: Gr+2012 =
       Greene et al. (2012); Ru+2014 = Ruiz et al. (2014); La+2012 = Lackner et al.
       (2012); Os+2012 = Oser et al. (2012); G\&D2012 = Gabor \& Dav{\'e} (2012); Hz+2013
       = Hilz et al. (2013); O\&H2013 = Oogi \& Habe (2013); B\&P2013 = B{\'e}dorf \&
       Portegies Zwart (2013).  
    }
    \label{figure:11}
    \end{figure*}

    \figurenum{11}

\subsection{Average Merger Mass Ratio}

    Finally, we use the inner-outer color difference and the slope of the $M_{\ast}$-color
    relations to estimate the luminosity-weighted average merger mass ratio that built the
    outer envelope.   Although the mass assembly history through mergers is a key
    ingredient for the theory of massive galaxy formation, it is extremely difficult to
    reconstruct using just the fossil records at $z\sim 0$: these mergers often involve
    galaxies in a range of mass ratio, morphology, and stellar population properties; and
    they happened across a very long time scale ($\sim 8-9$ Gyr).  Therefore, instead of
    getting into the details of the merging history, we propose to extract a key
    statistical value of these mergers--their average mass ratio--using photometric data
    alone, within the framework of the two-phase formation scenario.  The dense stellar
    systems formed during the first phase (through intense starburst induced by cold gas
    accretion and gas-rich mergers) become the inner core of nearby massive ETGs.  They
    quenched their star formation long ago through apparently very efficient processes,
    settled on the red sequence, and evolved along this sequence as they grew more massive
    through  mergers, all the while becoming redder (on average) due to passive evolution
    of the stellar population.  Based on theoretical predictions, most of these mergers
    should be non-dissipative, and, by number, minor mergers (mass ratio smaller than 1:3
    or 1:4) should dominate.  Such predictions have gained much recent observational
    support, such as the color statistics of satellites and tidal features around massive
    galaxies and the stellar population properties of the outer regions of nearby
    ellipticals.
    
    To achieve this goal, we need an empirical relation between the stellar mass and an
    observable property that satisfies at least the following two conditions: (1) the
    in-situ part of massive galaxies and most of the galaxies that merged into them follow
    this relation; (2) the shape or slope of the relation does not evolve strongly up to
    $z\approx 1$.  The red sequence of quiescent galaxies, or the $M_{\ast}$-color
    relation, is the best candidate we have.  As explained above, the non-dissipative
    nature of most mergers means that the less massive progenitors should be located close
    to the red sequence when the merger happened. Observations up to $z\approx 1$ suggest
    that, although the fraction of galaxies on the red sequence, the relative fraction of
    low- and high-$M_{\ast}$ galaxies, and the normalization  of the relation all show
    redshift evolution, its slope seems to be very stable (e.g., Mei\etal 2009; Fritz\etal
    2014; Cerulo\etal 2016) and does not depend on the environment.  These properties
    suggest that the evolution of the $M_{\ast}$-color relation is dominated by the aging
    of stellar population, making it appropriate for our application in this work.
    
    As shown in previous sections, the observed colors of the outer components are
    systematically bluer than those of the inner components, and are apparently ``too
    blue'' for the $M_{\ast}$ of the system (see left panel of Figure~9).  If the in-situ
    and accreted components are reasonably represented by our inner and outer components,
    the color offset we find reflects the difference in their average stellar population
    properties.  Using the assumptions discussed above, the color difference can be
    converted into the mass ratio between the stellar mass of the dense core ($M_{\ast,\rm
    core}$) and the {\it{typical mass of the satellite galaxy}}\/ ($\langle M_{\ast,\rm
    sat} \rangle$) using the $M_{\ast}$-color relations we derived:
  
      \begin{equation}
      \begin{aligned}
          \langle {\rm Mass\ Ratio} \rangle &= M_{\ast,\rm core} / \langle M_{\ast,\rm sat} \rangle \\
           &= 10^{({\langle{\rm Color}_{\rm inner}\rangle - \langle{\rm Color}_{\rm outer}\rangle})/\beta} ,
      \end{aligned}
      \end{equation}
   
    \noindent where $\beta$ is the slope of the adopted $M_{\ast}$-color relation.  The
    definition of mass ratio used here is strongly tied to our assumption of a two-phase
    formation scenario, and is similar to what has been used in other observational works
    (e.g., Arnold\etal 2011; Greene\etal 2012; Ruiz\etal 2014) and simulations (e.g.,
    Hilz\etal 2013; Oogi \& Habe 2013).  
    
    It is worth reminding the reader here that, in cosmological simulations, the average
    merger mass ratio is the number-weighted or mass-weighted mean of the instantaneous
    mass ratio of all the mergers, while under our assumption, we always compare the mass
    of the merging system with the in-situ core of the main progenitor. At the same time,
    we assume that all the stars assembled through mergers end up in our outer component.
    This picture becomes uncertain when major mergers occur, as the dense core of the
    slightly less massive progenitor may also become part of the inner component.  If that
    were indeed the case, our merger mass ratio should be seen as a lower limit.  (This
    issue will be revisited in the Discussion section.) Moreover, as our mass ratio
    estimate is based on the overall stellar population of the outer envelope, it should
    be considered a luminosity-weighted average, similar to the mass-weighted mean from
    simulations. 
  
    Starting with the color difference between the inner and outer components of the
    high-$M_{\ast}$ subsample, we use the best-fit slope of the $M_{\ast}$-color relations
    for CGS ETGs, along with their 1-$\sigma$ uncertainties, to estimate the average mass
    ratio and its associated uncertainty (blue and orange color bars on Figure~11).  For
    the median $\Delta(B-V)$, the mass ratio ranges between 8:1 and 12:1, while the range
    becomes 5:1 to 9:1 for the median $\Delta(B-R)$.  At the same time, the average mass
    ratio predicted by the SDSS samples are shown in Figure~11, too.  

    For the $M_{\ast}-(B-V)$ relation of the SDSS sample of ETGs defined by $P_{ {\rm
    E}+{\rm S0}} \gt 0.7$, its shallower slope results in a larger mass ratio $\sim$13:1.
    However, when a slightly different criterion is used to define the ETG sample
    ($P_{{\rm E}+{\rm S0}} \gt 0.5$), the slope of the $M_{\ast}-(B-V)$ relation is more
    consistent with that for CGS, resulting in a similar average mass ratio.  For the
    $B-R$ color, both SDSS samples predict a similar average mass ratio as CGS\@.  The
    dependence of the slope on the types of galaxies included in the mass-color relation
    reminds us another important aspect of our key assumption, which is that the
    mass-color relation used here should be an appropriate one for most galaxies involved
    in the assembly history of massive ETGs after $z\sim 1$.  Although more and more
    evidence suggests that non-dissipative mergers should dominate this process, possible
    contribution of disk galaxies also plays a role here, and will be discussed in detail
    in Section 5.2.

    As the $M_{\ast}$-color relation for $B-V$ has shallower slope than that for $B-R$,
    while the color difference is comparable [median $\Delta(B-V) \approx \Delta(B-R)
    \simeq 0.1$ mag for the high-mass sample], the $B-V$ color difference predicts a
    somewhat higher average mass ratio than $B-R$ (10:1 versus 5:1).  Considering that the
    average uncertainty of color is roughly the same for $B-V$ and $B-R$ ($R$-band data
    have slightly better signal-to-noise ratio for massive ETGs), the relation with
    shallower slope will naturally be more uncertain in predicting the mass ratio.  At the
    same time, since $B-R$ color is more sensitive to the change of stellar mass-to-light
    ratio of the underlying population (Bell \& de~Jong 2001), we expect the
    $M_{\ast}$-$(B-R)$ relation with a steeper slope to have better diagnostic power here.
    For these reasons, we favor the average mass ratio derived using $B-R$ color.
    However, such an offset is still a concern, as it may reflect systematic
    uncertainties.  It may also indicate the intrinsic limitation of optical colors for
    this application.  We will discuss this issue further in Section 5.3.

    Figure~11 compares our results with estimates from other recent observational and
    theoretical studies (Section~1), which define the average mass ratio as either
    number-weighted or $M_{\ast}$-weighted.  Because the average colors we are using are
    actually luminosity-weighted quantities, the derived average mass ratio using the
    $M_{\ast}$-color relation can no longer be seen as $M_{\ast}$-weighted.  Nevertheless,
    as both the inner core and outer envelope of massive ellipticals are dominated by a
    quiescent, old population, their average mass-to-light ratios are not significantly
    different.  When compared with other estimates, our results are still closer to the
    $M_{\ast}$-weighted ones than the number-weighted ones. In general, our estimates
    based on $B-R$ color are reasonably consistent with the average mass ratio predicted
    by recent simulations (Lackner\etal 2012; Oser\etal 2012; Gabor \& Dav{\'e}\etal 2012;
    B{\'e}dorf \& Portegies Zwart 2013; Hilz\etal 2013).  The high mass ratio suggested by
    Oogi \& Habe (2013) is number-weighted.  In Greene\etal (2012), the average merger
    ratio is derived in a similar spirit as ours (see their Figures~6 and 7), but based on
    the equivalent width of the Mg$_{\rm b}$ absorption feature and its relation with the
    velocity dispersion of the galaxy.  Their average ratio of $\sim$10:1 is consistent
    with our estimate based on $B-V$ color, but is higher than that based on $B-R$.

    The work of Ruiz\etal (2014) uses a very different strategy: the mass ratio statistics
    of satellites around nearby massive galaxies.  Smaller galaxies with 1:2 to 1:5 mass
    ratio dominate the stellar mass budget of satellites.  However, the authors concluded
    that when these satellites fall into their central host, ``the merger channel will be
    largely dominated by satellites with a mass ratio down to 1:10.'' Thus, this result
    also broadly agrees with ours, although we note that the satellite statistics at
    $z\approx 0$ may not give the full picture of all mergers that occurred during the
    last $\sim 8-9$ Gyr.  Of course, our analysis shares a similar uncertainty, as we are
    using the $M_{\ast}$-color relations at $z\approx 0$.  More recently,
    Rodriguez-Gomez\etal (2016) found that in Illustris (Vogelsberger\etal 2014)
    simulation, roughly 50\% of stars from mergers are contributed by mergers with mass
    ratio larger than 1:4; $\sim 25$\% come from mergers with mass ratio between 1:4 and
    1:10; the rest $\sim 25$\% are from even more minor mergers.  A naive estimate of
    ``mass-weighted'' merger mass ratio should be around 1:4 to 1:5, which is quite
    similar to our result using $B-R$ color.  As discussed later, it is possible that not
    all stars accreted through major mergers are located in our outer component, which
    will make two results even more consistent.

\section{Discussion}
    
    The mass assembly of massive galaxies is dominated by hierarchical merging processes
    in the last 8--10 Gyr.  Although it is not realistic for us to reconstruct their
    merging history just based on observations at low redshift, the statistical
    characteristics of such a process may provide useful constraints on the formation
    scenario.  In this work, we present a novel method to estimate the average mass ratio
    of mergers---an important parameter for the two-phase formation scenario---that built
    today's massive ETGs.  Although the negative color gradient of massive ETGs has been
    studied for decades, here we propose to explain it in a new context with the help of
    2-D modeling.  In light of the recent progress in understanding the evolution of
    massive galaxies, traditional interpretation of 1-D ``color gradients'' are too
    oversimplified.  Motivated by the ``two-phase'' formation scenario, we show that the
    multi-band photometric properties of nearby, massive ellipticals can be well described
    by the superposition of two sub-components with slightly different colors.  After
    investigating into the size, shape, scaling relations, and average mass profiles of
    these two components, we find that the outer components in our models can generally
    represent the accreted stars from all previous accretion events.  Assuming most
    mergers that happened to these massive ellipticals after $z\approx 2$ are
    non-dissipative, we use the $M_{\ast}$-color relation of quiescent galaxies to infer
    the mass ratio between the in-situ part of current massive galaxies and red galaxies
    that share the same average optical color with the accreted component.  We demonstrate
    that it is possible to extract crucial information regarding the assembly history of
    massive galaxies without getting into the complex details of stellar population
    models.  Our results are consistent with the expectation that minor mergers are more
    important for the build up of the extended envelope of massive ellipticals, and also
    consistent with studies based on different methods (e.g., photometric: La~Barbera\etal
    2012; spectroscopic: Greene\etal 2013).  All that said, we still need to emphasize
    here that our analysis alone cannot lift the degeneracy between the number of mergers
    and the average merger mass ratio.  We cannot provide more information about the
    merging history other than a luminosity-weighted average merger mass ratio.  The real
    merger mass ratio certainly does not have a normal distribution centered on such an
    average value (major mergers contribute more in mass, while minor ones contribute more
    in absolute number).

\subsection{Color Gradients for the In-situ and Accreted Components}
   
    The most important assumption in this study is that the color gradient arises from two
    different stellar populations, tracing the in-situ and accreted components, which can
    be separated spatially.  This is supported by recent simulations (e.g.,
    Hirschmann\etal 2015).  Although the extreme outer envelopes of nearby ellipticals
    sometimes exhibit decoupled kinematics and stellar population properties (e.g.,
    Coccato\etal 2010, 2013) the observed color gradient of these galaxies are mostly
    smooth and featureless, which makes them easily reproduced by superposition of two
    sub-components with constant color.  However, it does not rule out the possibility
    that both the in-situ and accreted components can have their own color (and underlying
    stellar population) gradients.  Formed through highly dissipative processes, the
    intrinsic color gradient of the in-situ (in our work, inner) component may tell us
    more about their formation history.  If no major merger has taken place, the
    relatively steep color gradient from intense star formation induced by efficient cold
    gas accretion should have been preserved to $z \approx 0$. However, one major merger
    alone may be able to erase such steep gradients.  And, for the accreted component,
    recent simulations (e.g., Rodriguez-Gomez\etal 2016) suggest that accreted stars from
    different mergers are redistributed to increasingly larger radii in the order of
    decreasing merger mass ratio.  This could naturally lead to color gradient in the
    accreted component. Of course, during moderate or high mass ratio mergers, a certain
    level of mixing must occur, which leads to smoother color gradient when the inner and
    outer parts of these galaxies are assembled.
    
    Although it is already possible to discuss these properties using simulations,
    observational confirmation is still extremely difficult.  In practice, we could give
    more freedom to both inner and outer components (e.g., free \ser index or effective
    radius) to simulate the color gradient of each component, but this will lead to more
    complicated internal degeneracies.  Meanwhile, if reliable photometry is available at
    the  low-surface brightness, outer regions of these galaxies ($> 4 R_{\mathrm{e}}$),
    it is still possible to study the color gradient of accreted stars alone under the
    assumption that in-situ stars make little contribution there.  However, this requires
    very accurate background subtraction in multiple bands, which makes the data reduction
    very challenging.
    
    Another issue is the definition of in-situ and accreted components.  In simulations,
    stars formed within the halo of the main progenitor (the more massive one) is often
    referred to as the in-situ part, while everything else is accreted.  In reality, we
    know that most massive ellipticals experienced major mergers, in which the ``less
    massive'' progenitor itself can be quite massive and have its own in-situ and accreted
    components.  In simulations of non-dissipative major mergers, the rank-order of
    binding energy (radius) is often preserved (Barnes 1988; Hopkins et al. 2009b), which
    means the in-situ (inner, dense) components of both progenitors will remain as the
    inner component of the final merger remnant.  Under this picture, our inner component
    may better represent the combined in-situ stars from all progenitors involved in past
    major mergers.  If this is true, it means that the color difference between the inner
    and outer components could bias the average merger mass ratio toward lower values.  It
    would be interesting to test this idea using simulations.
 
\subsection{Uncertainty of the Slope of Mass-Color Relation}
    
    As shown earlier, there is still uncertainty in the slope of the mass-color relation,
    and it directly impacts the merger mass ratio we measured.  Such uncertainty
    highlights another issue in our assumption: we should use the $M_{\ast}$-color
    relation followed by the in-situ components and most galaxies involved in the merger
    history.  Disk galaxies must also contribute to the build-up of stellar envelope of
    massive ellipticals, and they follow a steeper $M_{\ast}$-color relation.  Considering
    the distribution of their stellar mass, the inclusion of more disk galaxies makes the
    overall $M_{\ast}$-color relation steeper, and has larger scatter at the low-mass end.
    This changes the average merger mass ratio toward higher values.  Other than the
    S0-like, red disk galaxies we considered, there are other bluer, star-forming disk
    galaxies that live below the red sequence.  However, as the outer part of elliptical
    galaxies is not younger than the inner core, we think that the impact of star-forming
    disk galaxies in our analysis is negligible.
    
    Besides this uncertainty, we should also point out that systematic change of stellar
    metallicity is likely the main reason behind both the color gradient in each galaxy
    and the slope of the red sequence.  It is known that the relation between mass and
    stellar metallicity shows a larger scatter at the low mass end (Gallazzi\etal 2005;
    Panter\etal 2008; Gonz{\'a}lez Delgado\etal 2014).  Although the conversion between
    optical color and metallicity is complex, we cannot rule out the possibility that our
    $M_{\ast}$-color relation does not capture the low-metallicity, quiescent galaxies
    that are significant bluer than the red sequence.
    
    The merging history of massive galaxies can be intrinsically complex in nature, but we
    want to emphasize here that our goal is to study the luminosity-weighted average
    properties of all accreted stars, and we think that we can still safely assume that
    the average behavior of mergers in the last 8--9 Gyr can be approximated using
    galaxies on the red sequence.  It is certainly worth looking into these details using
    simulations and observations at higher redshift.

\subsection{Details about the Minor Mergers}

    Another assumption of our method is that dry, minor mergers dominate the mass growth
    of massive ellipticals in the second phase of their evolution.  To be more specific,
    we assume that most mergers are between a massive quiescent galaxy and its less
    massive satellites, and that by the time the merger occurred, the satellites have
    already become quiescent galaxies on the red sequence. This is a strong assumption,
    considering that these mergers span a large range in mass ratio and happened within a
    very long timescale (8--9 Gyr).  In general, such a scenario is supported by
    simulations, but we still need more observational constraints.  In the nearby
    Universe, merger statistics (e.g., Ruiz\etal 2014) do support the notion that minor
    mergers dominate the merging channel of massive galaxies.  Moreover, Ruiz\etal (2015)
    suggests that massive ellipticals tend to have more satellites, both in number and in
    stellar mass fraction, than galaxies of other morphologies.  At the same time, many
    different studies also highlight the importance of major mergers (or, at least,
    ``massive'' minor mergers) in shaping massive galaxies, especially the contribution to
    the total stellar mass of the accreted component, at both low and high redshifts
    (e.g., Ferreras\etal 2014; Kaviraj\etal 2014; Ruiz\etal 2014). Massive halos and their
    central galaxies tend to have complex merger histories that might not be well
    described by an average merger mass ratio.  Nevertheless, the average mass ratio is a
    convenient, if blunt, tool for comparison with simulations.  
  
    Regarding the notion that the mergers were non-dissipative, there are several lines of
    supporting evidence.  In the local Universe most tidal features around massive ETGs
    are consistent with an origin in dry mergers (e.g., Tal\etal 2009; Gu\etal 2013).
    Satellite statistics based on SDSS suggest that most satellites around massive
    galaxies are red, regardless of the environment (e.g., Hansen\etal 2009 for groups and
    clusters; Wang\etal 2014 for isolated bright galaxies).  At higher redshift, similar
    studies also have shown that mergers do not introduce a significant younger population
    of stars (e.g., Ferreras\etal 2014).  All these lines of evidence indeed support the
    idea that massive galaxies mostly grow along the red sequence by merging with galaxies
    that are already on the same sequence, as suggested by theoretical work (e.g.,
    Skelton\etal 2009).  However, just dry mergers alone may not be able to explain all
    the structural evolution and scaling relations of massive galaxies since $z \approx 1$
    (e.g., Sonnenfeld\etal 2014).  It seems that a small fraction of stars formed in
    accreted gas is still necessary.  However, the detailed impact of such a process on
    the color distribution is still not clear.
  
    Just knowing that most mergers are dry is not enough.  At the same mass ratio,
    accreting a compact elliptical galaxy may result in different stellar distribution
    compared with accreting an S0-like galaxy.  Hence, it is crucial to learn more about
    the morphology, structure, and stellar population of satellites involved in these
    mergers.  Greene\etal (2013), through detailed comparison of chemical abundances,
    concluded that the envelope of massive galaxies are more consistent with stars in
    quiescent disk galaxies.  Here, just using optical color, we actually see the hint of
    a similar story.   As we tried to fit the $M_{\ast}$-color relation using both CGS and
    SDSS data, we noticed that the slope of the red sequence becomes shallower when only
    ellipticals are included.  Such shallower slope would result in much smaller average
    merger ratio, hence leading to an unrealistically large number of minor mergers.  This
    tension disappears when some S0s are included in the modeling of the red sequence.  As
    mentioned at the end of Section~3, the more S0s are included in the definition of red
    sequence, the steeper the slope becomes, as the average color of S0s is bluer than
    that of ellipticals.  This argues that mergers between progenitors of massive
    ellipticals and quiescent disk galaxies (S0s) could be quite important.  In light of
    this uncertainty, our understanding of the merging history of massive ellipticals will
    benefit from better comparison of  the stellar population between S0s and the
    outskirts of massive ellipticals.  More detailed observations of on-going dry mergers
    at gradually higher redshift will certainly help too. 

\subsection{Future Directions and Applications} 

    With all the above uncertainties in mind, the method proposed here and the current
    results can be improved in several ways.  In this work, although the 2-D models in $B$
    and $R$ band are tied to the ones in the $V$ band with little freedom, they are still
    fit separately.  Simultaneous multi-band modeling should help us extract more reliable
    colors in more than two bands, while a forward-modeling approach using an
    MCMC-Bayesian method can further help understand the internal degeneracies among
    parameters.  In addition, using a large sample of massive galaxies with
    well-calibrated photometry can help us more accurately constrain the $M_{\ast}$-size
    relation.  It is also possible that using a single optical color is not the best way
    to define the average stellar population properties of different components.  We
    might, for instance, utilize the average SED of the inner and outer components based
    on multi-band data (e.g., similar to La~Barbera\etal 2012), but instead of using SED
    fitting to get the age and metallicity profiles, we can search along the red sequence
    for galaxies with similar SED\@.  All these approaches will allow us to understand the
    offset of our results based on different colors, and significantly improve the current
    constraint of average merger mass ratio.  A sample of $\sim 2000$ massive ($\log
    \,(M_{\ast}/M_{\odot}) > 11.0$) elliptical galaxies at $z < 0.3$ has been selected
    from the Hyper-Suprime Camera (HSC) survey for a similar analysis (S. Huang\etal, in
    preparation).  As the HSC survey has much better seeing and spatial resolution, and is
    at least 2.5 mag deeper ($i$ band) than CGS and SDSS, it is perfect for 2-D
    photometric modeling of massive galaxies.  This sample also includes $\sim 80$
    brightest cluster galaxies.  Crucial but untouched issues, like the effect of
    environment or the relations of color difference with other properties, can be
    addressed.
  
    Another avenue to explore is the low-level tidal features or remnants of ongoing dry
    mergers around nearby ellipticals, perhaps via ultra-deep imaging.  Such analysis
    helps us understand the distribution of mass ratio of dry mergers around massive
    ellipticals at $z \approx 0$, and provides useful constraints on galaxy formation
    through comparison with simulations or semi-analytic models.  In Gu\etal (2013),
    multi-component modeling of NGC~4889 revealed a new set of shells, which can be traced
    back to a recent minor dry merger.  The authors estimated the mass ratio of the merger
    using the optical color of the shells and the tight red sequence observed in the Coma
    cluster.  Using deeper images from the HSC survey, it is clear that recent merging
    events are more common around massive galaxies (Tanaka\etal, in preparation).  Careful
    color measurements of these tidal features should provide better statistics of recent
    mergers than previous works.  In the near future, surveys conducted with facilities
    such as the Large Synoptic Survey Telescope (LSST) will be even more helpful in this
    direction.
  
    Next, a self-consistent model for the formation of massive galaxies must be able to
    correctly predict the evolution of color gradients too.  Color gradients of massive
    galaxies at high-$z$ are becoming available (e.g., Hinkley\etal 2001; Ferreras\etal
    2009; Gargiulo\etal 2011; Guo\etal 2011).  If multi-band decomposition can be applied
    to rest-frame optical images of massive galaxies at increasingly higher redshift, we
    may be able to directly witness the gradual build-up of their outer envelope through
    minor mergers, along with the formation of their negative color gradients.   
  
    Finally, large integral-field spectroscopic surveys like MaNGA (Bundy\etal 2015) will
    provide direct evidence of stellar population differences between the inner and outer
    regions of nearby massive ellipticals.  More importantly, rough star formation
    histories can be inferred from comparison between data and synthetic stellar
    population models.  The photometric method proposed here can serve as an efficient and
    effective complement to current integral-field spectroscopic surveys.

\section{Summary} 

    We extend the multiple-component photometric decomposition of nearby elliptical
    galaxies in CGS into different filters.  We demonstrate that consistent
    three-component models can be obtained independently from images in the $B$, $V$, and
    $R$ bands, when the PSF and the sky background are properly taken into account.  Using
    reasonable parameter constraints, we extract $B-V$ and $B-R$ colors of different
    photometric components.  The combination of these components with slightly different
    color can perfectly recover the negative color gradient of massive elliptical
    galaxies, as well as their radial profiles of surface brightness and geometric
    parameters.  In the recent work by Huang\etal (2013b), these photometric components
    relate to physical structures that were formed through different mechanisms and at
    different phases of the evolution of massive galaxies.  To be more specific, the inner
    structure is similar to the massive, compact galaxies seen at $z \geq 1.5$, which most
    likely formed through a highly dissipative process.  On the other hand, the outer
    component can be best explained by the gradual accretion of stellar material through
    many, mostly dry, mergers.  In light of this evidence, we measured the average color
    difference between the inner and outer components for massive ellipticals in our
    sample ($M_{\ast} \geq 1.5 \times 10^{11} M_{\odot}$; $\langle\Delta(B-V)\rangle
    \approx 0.10$ mag; $\langle\Delta(B-R)\rangle \approx 0.14$ mag).  By adopting a few
    assumptions from the two-phase formation scenario, and by deriving the
    $M_{\ast}$-color relations for nearby ETGs, we translated these color differences into
    an average merger mass ratio for the build-up of the outer envelope.  According to our
    estimates, the average merger ratio lies between 1:5 to 1:10.  This is in line with
    the prediction that minor, dry mergers should dominate the growth of massive
    elliptical galaxies during the last few Gyr.  The method we proposed here only relies
    on decent photometric data in more than one filter, hence has great potential to be
    applied to existing (e.g., SDSS, CANDELS), ongoing (e.g., HSC Survey), and future
    (e.g., LSST) large-scale photometric surveys.  
  
  
\acknowledgements 
   
    We thank Andrew Cooper for generously providing data from his work to us for
    comparison.  This work was supported by Carnegie Observatories, Peking University, the
    Kavli Foundation, the Chinese Academy of Science through grant No. XDB09030102
    (Emergence of Cosmological Structures) from the Strategic Priority Research Program,
    the National Natural Science Foundation of China through grant No. 11473002 (LCH) and
    No. 11403072 (ZYL), the UC Irvine School of Physical Sciences (AJB), and the World
    Premier International Research Center Initiative, MEXT, Japan (SH). SH thanks Prof.
    Q.-S. Gu and the School of Space Science and Astronomy in Nanjing University for
    providing long-term support. ZYL is grateful for the support from the Shanghai Yangfan
    Research Grant (No. 14YF1407700). 

    Funding for SDSS-III has been provided by the Alfred P. Sloan Foundation, the
    Participating Institutions, the National Science Foundation, and the U.S.  Department
    of Energy. The SDSS-III web site is http://www.sdss3.org.  SDSS-III is managed by the
    Astrophysical Research Consortium for the Participating Institutions of the SDSS-III
    Collaboration including the University of Arizona, the Brazilian Participation Group,
    Brookhaven National Laboratory, University of Cambridge, University of Florida, the
    French Participation Group, the German Participation Group, the Instituto de
    Astrofisica de Canarias, the Michigan State/Notre Dame/JINA Participation Group, Johns
    Hopkins University, Lawrence Berkeley National Laboratory, Max Planck Institute for
    Astrophysics, New Mexico State University, New York University, Ohio State University,
    Pennsylvania State University, University of Portsmouth, Princeton University, the
    Spanish Participation Group, University of Tokyo, University of Utah, Vanderbilt
    University, University of Virginia, University of Washington, and Yale University.


{}


\appendix

\section{A. Comparison of color differences using 1-D color profiles}

    As mentioned in the main text, we suggest that the 2-D multi-component modeling method
    can help us gain more insight into the assembly history of massive galaxies through
    their 2-D color distribution.  In our method, while the detailed surface brightness
    distribution and 2-D geometric information are accurately recovered, the negative
    color gradient can also be well described by the superposition of two distinct
    components that have slightly different color.  As shown in Figure 2 (also see Figure
    10 of Rodriguez-Gomez\etal 2016), the accreted component (represented by our ``outer''
    component) and the in-situ component (``inner'' component) overlap substantially in
    radius: the accreted component starts to dominate over the in-situ component from 2
    $R_{\mathrm{e}}$ to 4 $R_{\mathrm{e}}$, while still contributing a significant
    fraction of stellar mass inside $R_{\mathrm{e}}$.  Hence, it is intrinsically
    difficult to extract the average color of both the in-situ and accreted components
    using just the 1-D color profiles.  However, as the 1-D color gradient is still
    commonly used, here we briefly compare the color difference between sub-components as
    derived from our 2-D method and the one derived from 1-D color profiles.  As shown in
    Figure 5, due to the strong impact from uncertainty of background subtraction, the
    color profile in low surface brightness regions typically has large uncertainties.
    Since we do need a large dynamical range in radius to estimate a meaningful color
    difference to compare with our results, we decide to use the 1-D color profiles based
    on the best, multi-component models instead of the observed one. 
    
    The PSF convolution is removed from the model to help us recover the intrinsic color
    at the center (although the radius range we use is not affected by the PSF).  And the
    same set of isophotes used to extract the 1-D surface brightness profile from data is
    applied to the model images in both bands (This is done using the ``forced
    photometry'' mode of the {\tt ELLIPSE} software.)  Compared to the color profiles
    derived from the original images, the ones from the best models do not show a sudden
    change at large radii.  For comparison, we use two methods to define the range of
    radius: (1) we use the color difference between 3 and 10 (20, 30) Kpc; (2) we estimate
    the color difference between $0.5$ and $2$ ($3$, $4$) times $R_{\mathrm{e}}$ (from
    single \ser model).  The results are summarized in the left and middle panel of Figure
    A1.  In general, there is a scattered correlation between these two methods, and the
    trends become clearer for larger dynamical range in radius.  But, as expected, the
    typical color difference from 1-D profile across all the radius range is smaller than
    the one between two sub-components.  For $B-V$ color, the median color difference
    between 3 and 10 (20, 30) Kpc is 0.04 (0.06, 0.07) mag; the corresponding values for
    $B-R$ color are 0.04 (0.07, 0.08) mag.  When relative radius is used, the color
    difference between 0.5 $R_{\mathrm{e}}$ and 2 (3, 4) $R_{\mathrm{e}}$ is 0.04 (0.07,
    0.07) mag for $B-V$, and 0.05 (0.07, 0.08) mag for $B-R$.  D'Souza et al.  (2014)
    provided median $g-r$ color profiles of massive, early-type (high concentration index)
    galaxies using stacked SDSS images.  The $g-r$ color can be approximately converted to
    $B-V$ color using empirical relations by Lupton et al. (2005).  Assuming $\Delta (B-V)
    = 0.891 \times \Delta (g-r)$, the expected $B-V$ color difference between 3 and 10
    (20, 30) Kpc is 0.05 (0.07, 0.08) mag.  Meanwhile, La~Barbera\etal (2012) provided
    median $g-r$ color profiles for massive galaxies scaled relative to $R_{\mathrm{e}}$.
    Using the same assumption, the expected $B-V$ color difference between 0.5
    $R_{\mathrm{e}}$ and 2 (3, 4) $R_{\mathrm{e}}$ is 0.04 (0.06, 0.07) mag.  Although
    these profiles were derived using different methods, they are quite consistent with
    our results.  This suggests that our multi-component models reliably capture the color
    profile in this radius range.  In the right panel of Figure A1, we also show the
    $M_{\ast}$-$\Delta (B-V)$ and $M_{\ast}$-$\Delta (B-R)$ relations in form similar to
    Figure 10.
    
\begin{figure*}[htbp]
    \centering 
       \includegraphics[width=18.5cm]{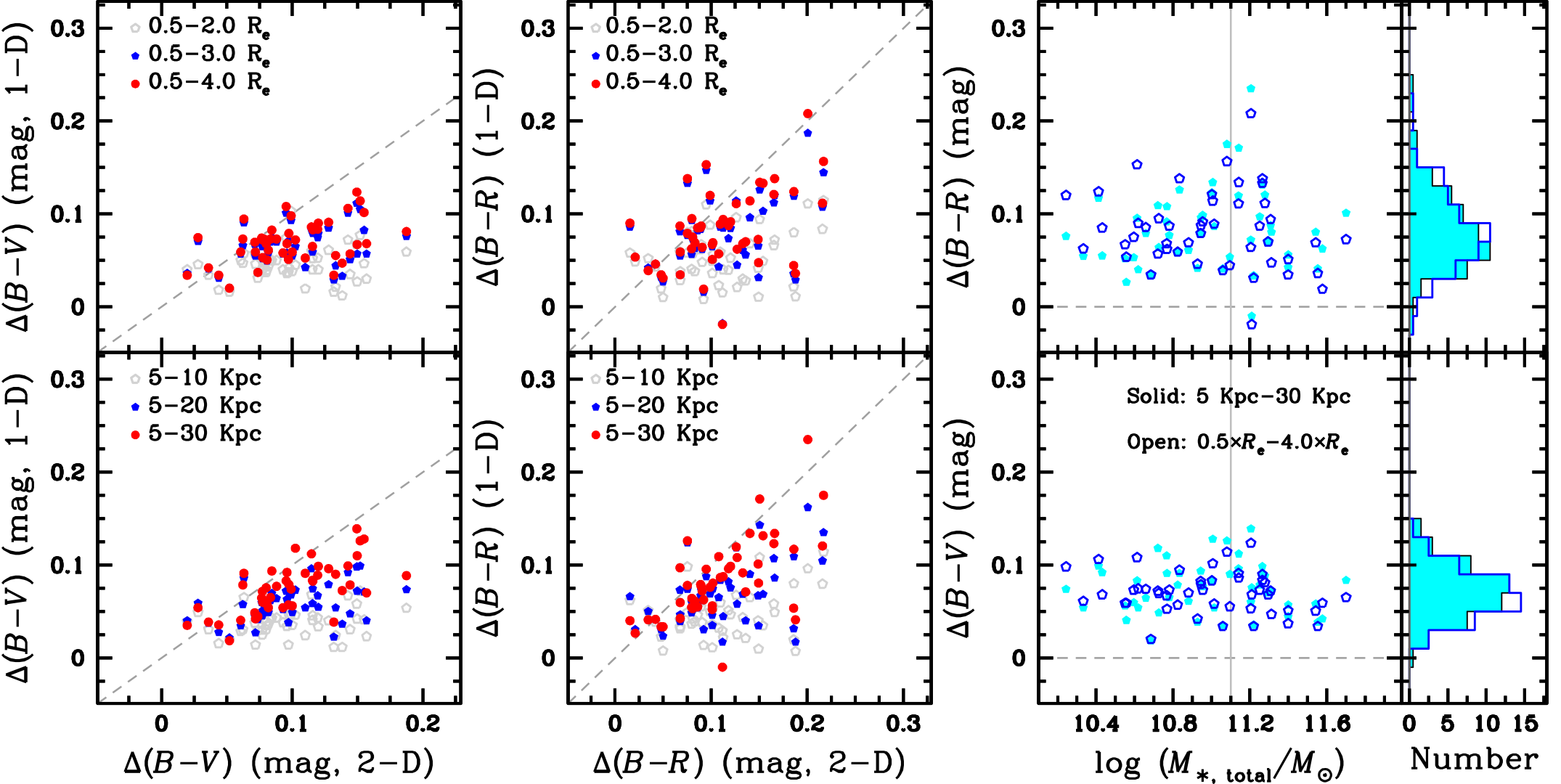}
       \caption{The left ($B-V$) and middle ($B-R$) panels compare the color difference
           between the inner and outer components with the color differences derived from
           1-D color profiles based on the best 2-D models.  On the lower panels, the
           color difference is defined by fixed absolute radius, while the upper ones use
           distance defined relative to effective radius.  A grey dashed line shows the
           one-to-one relation.  The right panel shows the relation between stellar mass
           and color difference derived from 1-D color profiles.
       }
    \label{figure:A1}
    \end{figure*}

    \figurenum{A1}

    \figurenum{TABLE 1}
    \begin{figure*}[htbp]
    \centering 
       \includegraphics[width=18.0cm]{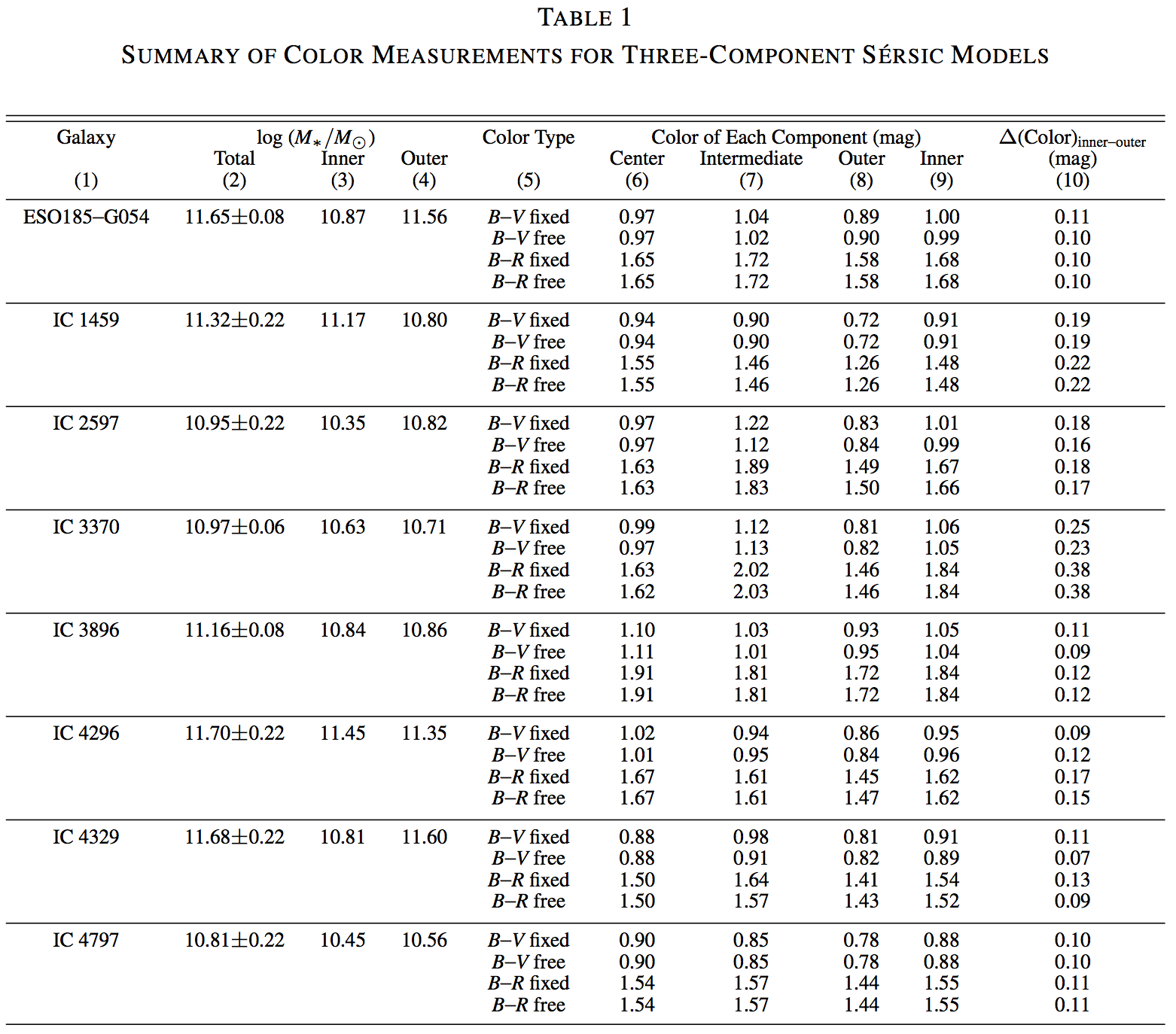}
       \caption{Summary of stellar mass estimations, the color of each component,
        and the color difference between the inner and outer components for Es included
        in this work.  For each galaxy, the $B-V$ and $B-R$ color information with fixed
        or free sky background are shown in different rows.  The free-background ones
        are recommended.    
        Col.   (1) Galaxy name.
        Col.   (2) Total stellar mass of the galaxy.
        Col.   (3) Stellar mass of the {\it inner} component.
        Col.   (4) Stellar mass of the {\it outer} component.
        Col.   (5) Type of color.
        Col.   (6) Color of the center component. 
        Col.   (7) Color of the intermediate component. 
        Col.   (8) Color of the {\it outer} component.  
        Col.   (9) Color of the {\it inner} component. 
        Col.  (10) The color difference between the inner and outer components. 
        The full catalog can be downloaded from 
        \texttt{https://github.com/dr-guangtou/cgs\_colorgrad/raw/master/table1.pdf}
        }
    \label{table:1}
    \end{figure*}


\end{CJK*}

\clearpage 

\end{document}